\documentclass[preprint,5p,times,twocolumn]{elsarticle}

\usepackage{amsmath,amsfonts}

\usepackage{array}
\usepackage{caption,subcaption}
\usepackage{textcomp}
\usepackage{stfloats}
\usepackage{url}
\usepackage{verbatim}
\usepackage{graphicx}
\usepackage{tabularx}
\usepackage{xcolor}
\usepackage{flushend}
\usepackage[linesnumbered,ruled,vlined]{algorithm2e}

\SetCommentSty{mycommfont}
\journal{Journal of Systems Architecture}
\begin{document}

% \title{UWB aided ray tracing for Digital Twin Approach to Wi-Fi 7 OBSS}
% \title{UWB aided ray tracing for Better Wi-Fi Spatial Reuse Through CCA and CBF}
% \title{UWB Assisted Digital Twin for Spatial Reuse in Wi-Fi 7 and Beyond}

\begin{frontmatter}

\title{Distributed Split Computing Using Diffusive Metrics for UAV Swarms}

\author[a,b]{Talip Tolga Sarı\corref{cor1}}
\ead{sarita@itu.edu.tr}

\author[a]{Gökhan Seçinti}
\ead{secinti@itu.edu.tr}

\author[b]{Angelo Trotta}
\ead{angelo.trotta5@unibo.it}

\cortext[cor1]{Corresponding author}

\address[a]{Department of Computer Engineering, Istanbul Technical University, Turkey}
\address[b]{Department of Computer Science and Engineering, University of Bologna, Italy}
% \address[c]{BTS - Digital Twin Application and Research Center, Istanbul Technical University, Turkey}

\begin{abstract}
In large-scale UAV swarms, dynamically executing machine learning tasks can pose significant challenges due to network volatility and the heterogeneous resource constraints of each UAV. 
Traditional approaches often rely on centralized orchestration to partition tasks among nodes. However, these methods struggle with communication bottlenecks, latency, and reliability when the swarm grows or the topology shifts rapidly. 
To overcome these limitations, we propose a fully distributed, diffusive metric-based approach for split computing in UAV swarms. Our solution introduces a new iterative measure, termed the \textit{aggregated computation capability}, capturing each node’s own computing capacity along with that of its neighbors without requiring global network knowledge. By forwarding partial inferences intelligently to underutilized nodes, we achieve improved task throughput, lower latency, and enhanced energy efficiency. Further, to handle sudden workload surges and rapidly changing node conditions, we incorporate a congestion-aware early-exit mechanism that can adapt the inference pathway on-the-fly. 
Extensive simulations demonstrate that our approach significantly outperforms baseline strategies across multiple performance indices, including latency, fairness, and energy consumption. These results highlight the feasibility of large-scale distributed intelligence in UAV swarms and provide a blueprint for deploying robust, scalable ML services in diverse aerial networks.

\end{abstract}


\end{frontmatter}

\section{Introduction}
The rapidly rising prominence of Unmanned Aerial Vehicle (UAV) swarms is transforming the way large-scale, time-critical missions are executed, enabling agile deployments across surveillance~\cite{ramachandran2021review}, disaster response~\cite{chandran2024multi}, smart agriculture~\cite{ammad2018uav}, and many other domains where ground infrastructure is limited or simply too slow to adapt~\cite{dai2022unmanned}. By orchestrating numerous autonomous aerial agents as a coordinated collective, UAV swarms can deliver compelling system-level benefits such as: wider and faster area coverage, built-in redundancy against individual failures, and superior adaptability to evolving mission objectives and environmental conditions~\cite{kurunathan2023machine, mohsan2023unmanned}. These very strengths, however, also make the problem substantially harder when one aims to deploy Machine Learning (ML) services at swarm scale. Approaches centered on a single controller or centralized server become increasingly fragile and inefficient as the swarm size and operational footprint grow~\cite{distributed_survey_1}. In practice, a centralized ground station can incur excessive communication overhead, suffer from latency bottlenecks, and introduce a single point of failure which is undesirable in exactly the high-mobility, high-stakes settings where UAV swarms shine. Furthermore, real-world swarms are inherently dynamic: nodes may join, leave, move, or reconfigure links unpredictably, and such churn amplifies the weaknesses of centralized paradigms while strengthening the case for distributed, scalable, and resilient ML service designs~\cite{islam2021survey}.

Recent efforts to alleviate centralized burdens typically advocate a form of distributed operation by distributing inference or training tasks across multiple nodes~\cite{distributed_survey_2}. However, they still often rely on global awareness or explicit synchronization. At smaller scales or in constrained network topologies, these partial-distribution schemes might suffice. Yet, as UAV swarms expand to hundreds or even thousands of nodes, it becomes infeasible to gather and coordinate global network information~\cite{Tahir2019SwarmsOU}. Additionally, heterogeneous node capabilities such as differences in on-board processing, sensing, and power resources demand an approach that can dynamically adjust to varying node states in near real-time. A node endowed with high computation power at a time may deplete its battery and become less capable a moment later, or an agile UAV that was ideally placed to handle tasks at one moment could relocate out of range in the next. These variations highlight the necessity for a self-organizing approach that pushes intelligence and decision-making to the edges.

Split learning and computing is particularly well-suited to address these challenges in large-scale UAV swarms~\cite{distributed_survey_2}. Rather than transmitting raw data or entire models to a single aggregator, split learning and computing partitions the model into layers distributed among various nodes, thereby enabling partial inference at local levels. By transferring only the necessary, intermediate representations, nodes can reduce communication overhead while maintaining data privacy. However, effectively implementing split computing in UAV networks requires strategic decision-making about when and where to forward intermediate layer  outputs~\cite{distributed_survey_1,distributed_survey_2}. This is where centralized schemes fail to perform well for large networks. They attempt to collect global knowledge of the network’s topology, node resource levels, and data distributions which is impractical in large, dynamic swarms. As a result, model splits and partial inferences often end up bottlenecked by nodes that unknowingly become overloaded or unreachable.
\begin{figure*}[htb]
    \centering
    \includegraphics[width=0.99\textwidth]{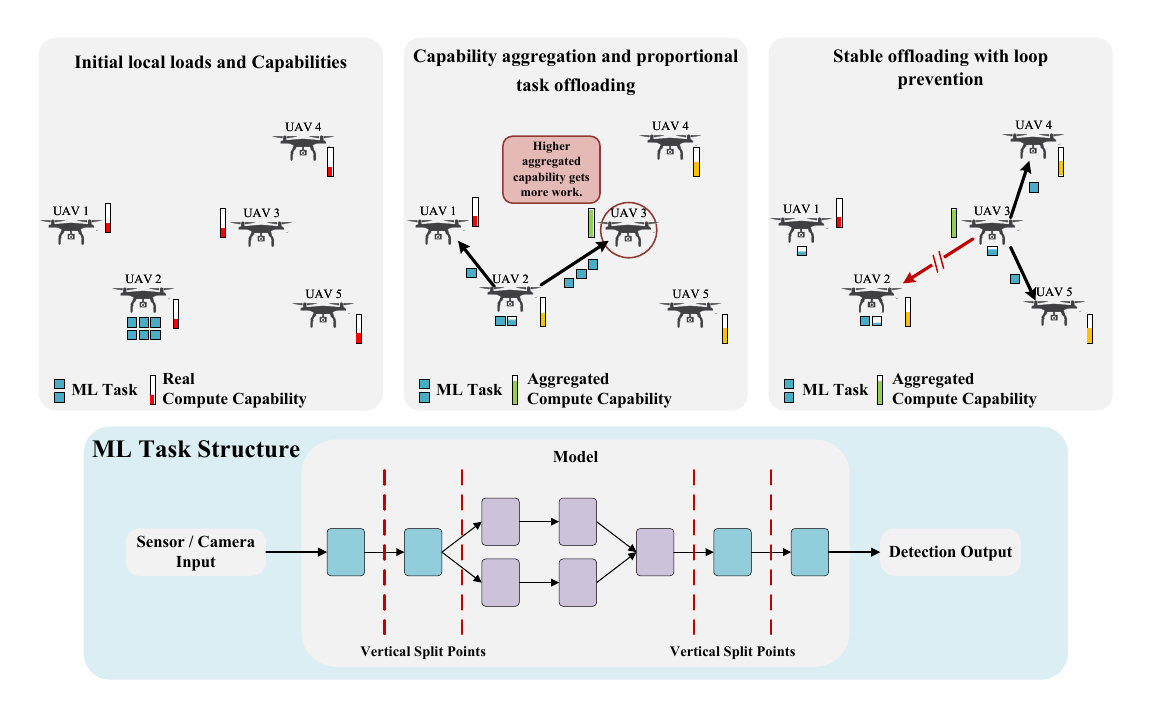}
    \vspace{-2em}
    \caption{Overview of the proposed distributed split-computing mechanism driven by diffusive aggregated computation capability: (left) Initial distribution of queued tasks and real \textbf{local computation capability} across UAVs. (middle) \textbf{Aggregated computation capability} propagates capability information through one-hop neighborhoods, and tasks are offloaded proportionally toward nodes with higher aggregated capability, even when their local resources are limited. (right) Loop-prevention yields a stable offloading pattern that converges without oscillatory task transfers. \color{black} The lower panel depicts an illustrative ML task structure for event-triggered aerial object detection (e.g., survivor sighting during a search-and-rescue mission), where bursty inference loads are distributed across the swarm. Vertical split points are placed only at layer boundaries where exactly one activation tensor is transferred between consecutive layers. Sequential blocks (blue) satisfy this condition at every internal boundary. Multi-branch blocks (purple) carry multiple concurrent tensors across parallel branches, preventing a vertical split, that has complete features for full inference, until the branches merge back into a single tensor.}
    \label{fig:scheme}
\end{figure*}
To overcome these limitations, we propose a fully distributed, diffusive metric-based approach in which each node autonomously assesses its own capabilities as well as those of its neighbors without relying on global information. The core idea is to estimate the “aggregated computation capability” of a node in a manner analogous to graph centrality metrics. In particular, our approach treats each node’s local computational power as well as its neighbors’ powers in an iterative fashion: if a node’s neighbors can handle substantial partial inferences, or if they hold a topologically advantageous position to further distribute the load, the node is more inclined to offload its tasks there. By repeating local updates, the entire network reaches a stable state of partial inference routing in which tasks flow toward underutilized, resource-rich parts of the swarm. This distributed decision-making ultimately fosters better throughput, lower latency, and more balanced energy consumption. Our method thereby aligns with the inherent nature of UAV swarms: distributed, dynamic, and operating over sprawling distances that defeat naive centralized solutions. \color{black} Figure~\ref{fig:scheme} provides an overview of our proposed distributed split computing framework within a large and dynamic UAV network. 
In the figure, consider \textit{UAV 2} capturing an image via its onboard camera. If its instantaneous computational capability is insufficient to process the full neural network within the required latency constraint, it offloads part of the split model execution to neighboring UAVs. The offloading decision is guided by the aggregated computation capability metric, described in the following sections, which accounts for both local resources and neighborhood contributions. In this way, inference is completed collaboratively without relying on centralized coordination.
\color{black}

To further enhance the system's adaptability under fluctuating workloads, we incorporate a congestion-aware early-exit mechanism that dynamically reduces the depth of inference based on real-time load conditions. This strategy allows nodes to truncate the model execution at predefined intermediate layers, producing acceptable outputs while avoiding the full computational cost of processing the entire model. By adjusting the exit point according to the current resource availability and task queue length, the system minimizes latency and prevents resource saturation. This approach aligns with the principles of early-exiting~\cite{matsubara2023split} where intermediate outputs offer a controllable trade-off between computational effort and inference accuracy, making it particularly suitable for dynamic and resource-constrained UAV environments.

By merging the advantages of split computing with a diffusive, graph-oriented update rule for computational metrics, our approach maintains robust inference quality across a large and rapidly evolving topology. It enables UAV swarms to shift inferences efficiently among nodes, ensuring that tasks do not pile up in isolated regions while idle capacity elsewhere remains unused. In contrast with conventional centralized strategies, which suffer from communication bottlenecks and single points of failure, our framework supports agility and scalability by design.

In this paper, our key contributions are as follows:
\begin{itemize}
    \item We identify the fundamental scalability challenges in deploying split computing across large-scale, dynamic UAV swarms, highlighting why centralized solutions are prone to failure in real-world scenarios.
    \item We introduce a novel metric, the \emph{aggregated computation capability}, to capture how a node’s computational load should be shared in conjunction with its neighbors’ resources. This metric is designed to be iteratively updated at each node, enabling a fully distributed and lightweight approach.
    \item We propose a new decision-making framework for routing partial inferences (either intermediate feature maps or raw input data) based on the aggregated computation capability metric. This framework avoids reliance on global synchronization, making it highly scalable and robust to node mobility or failure.
    \item We design a congestion-aware early-exit mechanism that dynamically truncates model execution based on local load conditions, reducing latency and preserving system stability under high workload scenarios.
    \item We conduct extensive simulations to verify that our distributed solution outperforms existing benchmark strategies across several metrics including inference performance, energy efficiency, fairness, and overall throughput, particularly in large UAV swarms.
\end{itemize}

 %The remainder of this paper delves into the details of the proposed distributed split learning architecture, covering background on current methods, our specific problem formulation, the new metric design, and the results of our thorough performance evaluation.
The remainder of this paper is organized as follows. Section~\ref{sec:related_work} reviews related work on distributed inference and task offloading in UAV networks. Section~\ref{sec:problem_formulation} formalizes the system model and problem definition, introducing the proposed diffusive metric and task allocation strategy. Section~\ref{sec:performance_evaluation} presents the simulation setup and provides a detailed performance evaluation under various operating conditions. Finally, Section~\ref{sec:conclusion} concludes the paper and outlines directions for future work.

\section{Related Work}
\label{sec:related_work}

% Preamble:
% \usepackage{booktabs}
% \usepackage{tabularx}

% Preamble:
% \usepackage{booktabs}
% \usepackage{tabularx}
\begin{table*}[t]

\centering
\caption{Comparison of related works against our fully distributed approach.}
\label{tab:lit_review}
\scriptsize
\setlength{\tabcolsep}{3.5pt}
\renewcommand{\arraystretch}{1.12}

\begin{tabular}{p{1.2cm} p{2.4cm} p{2.5cm} p{1.8cm} p{1.8cm} p{1.8cm} p{3.1cm}}
\hline
Work & Control Architecture & Required State & Mobility & Adaptability & Scalability & Use case \\
\hline

\cite{modnn}
& Centralized
& Global
& Low
& Low & Low-Medium
& Split computing \\

\cite{disnet}
& Centralized
& Global
& Low
& Low & Medium
& Split computing \\

\cite{Cao2024Multitier}
& Hierarchical
& Multi-tier
& Low
& Medium & Low-Medium
& Split computing \\

\cite{zeng2020}
& Centralized
& Global
& Medium
& Medium-High & Medium
& Split computing \\

\cite{lin2024epsl}
& Centralized
& Global
& Low
& Low-Medium & Low-Medium
& Split learning \\

\cite{kim2024edgemsl}
& Centralized
& Global
& Medium
& Medium & Medium
& Split learning \\

\cite{hybridsplit}
& Centralized
& Global
& Medium
& Medium & Medium
& Split learning \\

\cite{wu2023}
& Centralized
& Cluster
& Low-Medium
& Low-Medium & Low-Medium
& Split learning \\

\cite{yeom2024uav}
& Centralized
& Global
& High
& Low-Medium & Low-Medium
& Split computing\\

\cite{fei}
& Centralized
& Global
& High
& Low-Medium & Low-Medium
& MEC task offloading \\

\cite{cloud_mec}
& Hierarchical
& Multi-tier
& Low
& Low & Low-Medium
& MEC task offloading \\

\cite{zhou2023dag}
& Centralized
& Global
& Low-Medium
& Low-Medium & Low-Medium
& MEC task offloading \\

\cite{zheng2024multi}
& Distributed
& Global
& High
& Medium-High & Medium
& Joint trajectory/offloading \\

\cite{focus}
& Hierarchical
& Multi-tier
& High
& Low-Medium & Medium
& Joint trajectory/offloading  \\

\cite{agarwal2024federated}
& Centralized
& Global
& High
& Low-Medium & Low-Medium
& MEC task offloading \\

\textbf{This work}
& \textbf{Distributed}
& \textbf{1-hop neighbors}
& \textbf{High}
& \textbf{High} & \textbf{High}
& \textbf{Split computing} \\
\hline

\end{tabular}

\vspace{4pt}
\footnotesize\textit{\textbf{Control Architecture} indicates where orchestration/scheduling decisions are made; \textbf{Required State} denotes the minimum state visibility needed to operate; \textbf{Mobility} captures the assumed node mobility level; \textbf{Adaptability} reflects responsiveness to topology/compute changes with bounded coordination; \textbf{Scalability} summarizes how overhead and state requirements grow with network size.}
\end{table*}

Recent research has explored distributed Deep Neural Network (DNN) inference and task offloading to address the challenges posed by limited resources and dynamic network conditions. Table~\ref{tab:lit_review} compares representative studies with our fully distributed approach. The key distinction is that many prior methods depend on centralized control and/or global optimization, which requires global measurements and coordination, which makes them slow to track fast topology and link dynamics in UAV networks.

In~\cite{modnn}, MoDNN partitions DNN computations among mobile devices to balance workload and reduce communication overhead. Its model partitioning strategy minimizes transmission delays while exploiting parallel processing, making it highly applicable to resource-constrained environments, yet the coordination follows a centralized control pattern (as designated coordinator that schedules execution and aggregates intermediate results), which requires shared system state beyond purely local observations. Similarly, DISNET~\cite{disnet} introduces a micro-split framework that partitions deep neural network models both vertically and horizontally, enabling cooperative inference among heterogeneous Internet of Things (IoT) devices while reducing latency and energy consumption through adaptive workload allocation. However, it relies on network-wide resource/topology awareness for orchestrating the split.

On the split computing front,~\cite{Cao2024Multitier} propose a multi tier framework that employs deep reinforcement learning to optimize model splitting and resource allocation in User Equipment (UE)–Edge–Cloud systems, decomposing complex optimization problems into sequential subproblems to balance computational and communication loads. In a similar vein, CoEdge~\cite{zeng2020} demonstrates that dynamic workload partitioning, formulated as a constrained optimization problem, can yield substantial latency speedups, with a fast approximation algorithm that adapts in real time to varying device capabilities and network conditions. Similarly for Split Learning,~\cite{lin2024epsl} develop an efficient parallel split learning (EPSL) framework for resource-constrained wireless edge networks. Their approach jointly optimizes subchannel allocation, transmit power, and model split point selection to minimize latency and communication overhead.~\cite{kim2024edgemsl} propose Edge-MSL, a split learning framework for mobile edge networks that leverages a contextual multi-armed bandit formulation to dynamically select the optimal split point and assign edge servers. By adapting the split location and offloading decisions based on real-time network conditions, their approach significantly reduces training latency and enhances convergence. On the other hand,~\cite{hybridsplit} combines split learning with federated learning in UAV networks to address the trade-off between communication overhead and computational constraints, resulting in an architecture that achieves higher inference accuracy than standalone federated learning while reducing transmission costs compared to conventional split learning. Moreover,~\cite{wu2023} propose a parallel split learning framework over wireless networks that integrates dynamic resource management with parallel task offloading. Their method exploits concurrent processing across heterogeneous edge devices to substantially reduce training latency while maintaining balanced workload distribution.~\cite{yeom2024uav} investigate a UAV-assisted split computing system where a UAV acts as a mobile edge server to support collaborative inference among IoT devices. By partitioning a DNN between on-device processing and UAV-based computation, their design reduces transmission delays and energy consumption.

Task offloading approaches can also be useful for split learning approaches. For instance,~\cite{fei} develops a UAV-based environmental monitoring system that employs on-board processing to minimize transmission delays and maintain low data freshness latency, thereby enhancing overall system capacity in remote areas.~\cite{cloud_mec} proposes a cloud Mobile Edge Computing (cloud-MEC), offloading scheme that integrates differentiated offloading decisions with service orchestration via an ODaS mechanism, significantly reducing both delay and energy consumption by balancing computational and communication loads.~\cite{zhou2023dag} addresses the challenge of offloading dependent tasks modeled as directed acyclic graphs using a branch soft actor-critic algorithm, which optimizes offloading, computation, and cooperation decisions under dynamic conditions to reduce latency and improve energy efficiency. More recently,~\cite{zheng2024multi} presents a multi-agent collaborative framework that jointly optimizes UAV trajectory planning and Directed Acyclic Graph (DAG) task offloading via deep reinforcement learning, with their TD3-TT algorithm demonstrating further latency reductions and enhanced real-time performance. The FOCUS framework~\cite{focus} leverages fog computing in UAV networks by jointly optimizing routing and computation. Its central SDN-based architecture and analytical approach achieve significant latency improvements by effectively integrating on-board processing with data forwarding. Lastly,~\cite{agarwal2024federated} investigate task offloading in a UAV-assisted MEC network by leveraging a federated learning framework. In their work, a UAV equipped with MEC capabilities serves as a mobile server to assist resource-constrained UE in offloading computation-intensive tasks.

In summary, existing solutions often assume a centralized controller or a global optimization loop that repeatedly aggregates network-wide information before updating decisions. While such designs can be effective in stable/semi-stable settings, their reliance on global information acquisition and coordination increases the adaptation latency. As a result they cannot have timely reaction to rapid topology or compute changes. In contrast, our approach is fully distributed and eliminates global optimization and centralized decision-making; nodes update decisions using local/neighbor information, enabling faster topology tracking while keeping signaling overhead bounded.

\section{Problem Formulation and System Model}
\label{sec:problem_formulation}

The core challenge lies in determining how and where to execute computation tasks to achieve low latency, balanced resource utilization, and robust performance, especially when the network topology changes frequently and UAVs have heterogeneous capabilities. We assume that the full neural network model, along with its corresponding early-exit parts, is pre-trained and deployed on the UAVs prior to operation. This ensures that each node can execute both complete and truncated inferences locally, according to its current workload and resource availability.
To address these challenges, our formulation focuses on four interlinked aspects: (i) modeling communications among UAVs, (ii) defining a distributed metric for aggregated computation capability, (iii) making task-transfer decisions based on local utilization, and (iv) applying a congestion-aware early-exit mechanism to adapt to real-time workload fluctuations. The following subsections describe each component in detail.

\subsection{Task Computation Model}

We consider machine learning tasks that are partitioned into L layers, represented as nodes in a DAG for the Machine learning task. The execution order of this task $T$ is decided by this DAG. Each layer $l \in \{1,2,\ldots,L\}$ requires $G_l$ GFLOPs ($10^9$ floating point operations per second) to complete. Each UAV $v_i$ provides a computational capacity of $F_i$ GFLOPs per second. Thus, the delay to compute layer $l$ at UAV $v_i$ is given by:
\begin{equation}
    d^{\mathrm{comp}}_{l, i} = \frac{G_l}{F_i}
\end{equation}
When processed entirely on a single node $v_i$, the overall computation delay is determined by the sum of delays along the longest dependency chain in the DAG. For tasks with purely sequential dependencies, this simplifies to
\begin{equation}
    d^{\mathrm{comp}}_{T, i} = \sum_{l=1}^{L} d^{\mathrm{comp}}_{l, i}
\end{equation}

In scenarios where intermediate layer outputs are offloaded during processing, if the offloading decision occurs before the completion of a computation, any partially computed work is discarded.

\subsection{Communication Model}
We consider a UAV network in which nodes are distributed across a region and follow circular trajectories.  Moreover, connectivity varies as time evolves. At time $t$, UAV $v_j$ is considered a one-hop neighbor of $v_i$ if the received link quality satisfies $\mathrm{SNR}_{ij}(t)\geq \mathrm{SNR}_{\min}$.
For each feasible link $(v_i \rightarrow v_j)$, we approximate the instantaneous link speed by the Shannon capacity upper bound, and we use this value to compute transmission latency of offloaded intermediate activations.
Accordingly, the channel capacity between nodes $v_i$ and $v_j$ is computed as~\cite{shannon}:

\begin{equation}
    C_{ij}(t) = B\log_{2}\!\left(1 + 10^{\mathrm{SNR}_{ij}(t)/10}\right)
\end{equation}

The term $\mathrm{SNR}_{ij}(t)$ is the received signal-to-noise ratio (in dB) for transmissions from node $v_i$ to node $v_j$ at time $t$, and $B$ is the channel bandwidth in Hertz. We use two-ray ground-reflection model to calculate pathloss, where the receiver is assumed to receive a line-of-sight component and a single ground-reflected component~\cite{rappaport2010wireless}. Then, the SNR for transmission from node $v_i$ to node $v_j$ can be written as:

\begin{equation}
    \mathrm{SNR}_{ij}(t) = P_{i} + L(v_i,v_j,t) - N_0
\end{equation}

Here, $P_i$ is the transmit power of node $v_i$ (in dBm) and $N_0$ is the thermal noise power (in dBm). $L(v_i,v_j,t)$ denotes the two-ray pathloss for the pair $(v_i,v_j)$ at time $t$.

When UAV $v_i$ offloads a partially processed task to a neighbor $v_j$, it transmits the intermediate activation produced at the current split point. Thus, if $S_l$ denotes the data size of the activation after layer/split $l$. We model the one-hop transmission latency as:
\begin{equation}
    d^{tx}_{i,j}  = \frac{S_l}{C_{ij}(t)}
\end{equation}
and this latency is added to the task’s end-to-end completion time. In addition, to reflect a practical radio constraint, we assume each UAV can maintain at most one ongoing task transfer at a time, while local computation continues in parallel.

\subsection{Energy Consumption Model}
For energy consumption, we model that every GFLOP computed consumes \(J\) joules. Therefore, the energy required to compute unit \(l\) is
\begin{equation}
    E_l = G_l \times J
\end{equation}
and the total energy consumption for a task processed entirely on one node is
\begin{equation}
    E_T = \sum_{l=1}^{L} G_l \times J
\end{equation}
Additionally, when offloading a computational unit from one UAV to another, the transmission energy consumption is given by:
\begin{equation}
    E^{tx}_l (t) = \frac{S_l}{C_{ij}(t)} \times P_i
\end{equation}

where $S_l$ denotes the data size of output of layer $l$, $C_{ij}$ is the channel capacity between nodes $v_i$ and $v_j$, and $P_i$ is the transmission power of node $v_i$ in Watts. This expression represents the product of the transmission time $\left(\frac{S_l}{C_{ij}(t)}\right)$ and the power used during transmission.

\section{Proposed Distributed Task Allocation and Congestion-Aware Early-Exit}

We adopt a distributed approach for UAVs because the network spans a large geographic area with dynamic UAVs. A centralized controller would need frequent global state collection and dissemination, which creates high messaging overhead and it quickly becomes outdated when topology changes. Our approach stays adaptive with only local coordination as each UAV shares lightweight hints with its one-hop neighbors, then these hints also diffuse through neighbor-based updates to guide task allocation without any global view. When local balancing is still not enough under sudden workload spikes, we trigger a congestion-aware early-exit only when necessary to reduce inference cost, prevent queue build-up, and avoid collapse while keeping full inference whenever resources permit.

\subsection{Aggregated Computation Capability}
Our decision-making focuses on offloading the current computation to the least utilized node and the utilization is determined by the \emph{aggregated computation capability} ($\phi$) of nodes, considering the resources of neighboring nodes.

In other words, the \emph{aggregated computation capability} can be viewed as a resource-aware centrality score, where classical centrality metrics (such as degree, betweenness and eigenvector) quantify how well-positioned a node is based purely on network topology. On the other hand our score reflects how effective a node would be for task allocation when both \emph{neighbor influence} and \emph{system resources} (compute and link delay) matter.

Nodes with powerful neighbors or central positions are more advantageous for task allocation. To identify such nodes, we aim to utilize graph centrality measures. Metrics like eigenvector centrality, which propagate importance through neighbors, appear suitable for capturing this effect. 
Eigenvector style centrality metrics are particularly relevant here because they are \emph{neighbor-recursive} meaning that a node becomes important because it is connected to other important nodes~\cite{eigenvector,pagerank}.
However, computing eigenvector centrality requires global normalization at every iteration, making it impractical for a fully distributed and rapidly changing UAV network.

To avoid global coordination while preserving the same neighbor-recursive intuition, we replace eigenvector centrality with the \emph{aggregated computation capability}. 
Unlike classic centralities, $\phi$ is interpreted as an effective processing rate (in terms of GFLOPs) under local load sharing, and it explicitly incorporates communication delay and computation capabilities. Moreover, unlike eigenvector centrality, it can be updated using only one-hop information without any global normalization.
In essence, each node updates its effective computation capability using only its own resources and the aggregated capability of its one-hop neighbors. At time $t$, we model the swarm as a time-varying graph and define the one-hop neighbor set of node $v_i$ as:
\begin{equation}
M_i(t) \triangleq \left\{\, j \;:\; j\neq i,\ \mathrm{SNR}_{ij}(t) \ge \mathrm{SNR}_{\min} \,\right\},
\end{equation}
where $\mathrm{SNR}_{\min}$ is the minimum SNR threshold required for communication. 

Given $M_i(t)$, we compute the aggregated computation capability iteratively. 
The intuition is that when a node splits an incoming load evenly among itself and its neighbors, the total completion time is dictated by the slowest participant. This leads to the following neighbor-recursive update for the aggregated computation capability $\phi_{i,t+1}$ of node $v_i$ at time $t+1$:
\begin{equation}
    \frac{1}{\phi_{i,t+1}} = \frac{1}{|M_i(t)| + 1}\big(\frac{1}{F_i} + \max_{k \in M_i(t)} \{d^{tx}_{i,k}(t) + \frac{1}{{\phi_{k,t}}}\}\big)
    \label{eq:agg_capability}
\end{equation}
\color{black}
where $F_i$ is the computation capability of node $i$ (in GFLOPs), $d^{tx}_{i,k}(t)$ is the average transmission delay per unit share of workload between nodes $i$ and $k$ (in seconds per GFLOP), and $\phi_{i,t}$ is the aggregated computation capability of node $i$ at time $t$ (in GFLOPs). Since $1/F_i$ and $1/\phi_{k,t}$ both carry units of seconds per GFLOP, and $d^{tx}_{i,k}(t)$ is expressed in the same units, all terms in the reciprocal are dimensionally commensurate. The aggregated computation capability is updated at every decision epoch $\Delta t$ using only one-hop neighbor information.
\color{black}
% where $F_i$ is computation capability of node $i$, $d^{tx}_{i,k}(t)$ is the average transmission delay between nodes $i$ and $k$, and $\phi_{i,t}$ is the aggregated computation capability of node $i$ and time $t$. The aggregated computation capability assumes that a given amount of work is divided evenly among the node and its neighbors ($|M_i(t)| + 1$). 
\color{black}
Since all participants work concurrently, the overall completion time is determined by the slowest node, whose aggregated capability is captured by the maximum term.
Because this update requires only the neighbors of the nodes, it naturally adapts to dynamic network conditions without requiring any global view. Thus, aggregated computation capability is a fully distributed, one-hop-update metric that jointly captures computation resources and link delay for task allocation decisions.

\color{black}
The iterative update in Eq.~(\ref{eq:agg_capability}) converges under any connected snapshot topology. To see this, observe that $\phi_i$ is naturally bounded: it is strictly positive because all computation rates $F_i$ and link capacities are positive and finite, and it cannot exceed the combined raw computation rate of a node and its neighbors, since nonzero transmission delays always reduce the effective collaborative throughput. Moreover, at each iteration the $\frac{1}{|M_i(t)|+1}$ averaging factor contracts the difference between successive estimates by at least a factor of two for every node with at least one neighbor ($|M_i| \geq 1$). Consequently, the residual change between iterations decreases geometrically, and the metric settles to a unique stable value within a small number of update rounds. In practice, the topology is time-varying; however, since convergence occurs within very few iterations and the decision period $\Delta t$ is short relative to the rate of topology change, the metric effectively tracks the current network state.

\color{black}

\begin{figure*}
    \centering
    \includegraphics[width=0.99\textwidth]{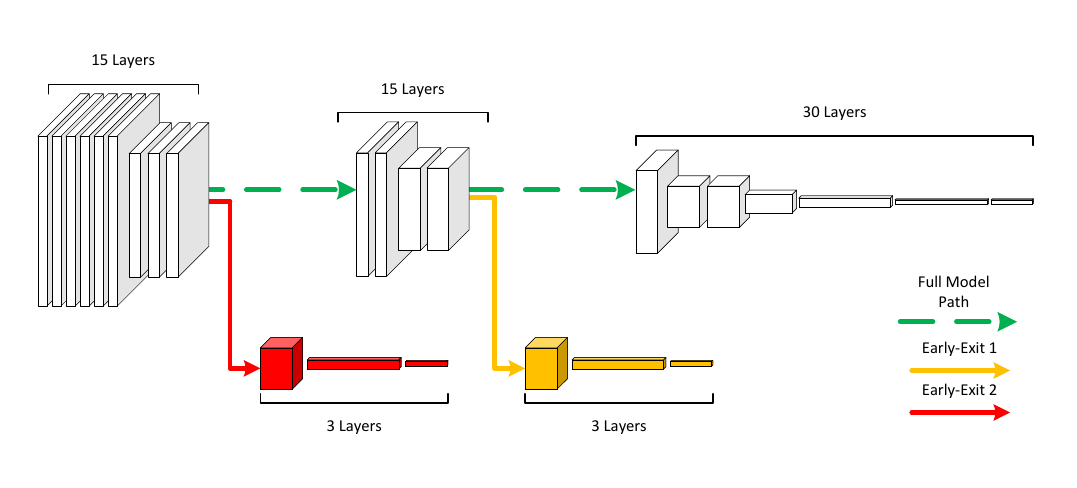}
    \vspace{-2em}
    \caption{Proposed congestion-aware early-exit mechanism.}
    \vspace{-1em}
    \label{fig:arch}
\end{figure*}
\subsection{Task Transfer Decision}

Our framework requires each UAV to make autonomous decisions about whether to process a task locally or transfer it to a neighboring node. These decisions are based on comparative utilization levels between nodes, mediated by a threshold parameter $\gamma$ that controls the sensitivity of task offloading. The utilization of node $i$ at time $t$ is calculated as the ratio of its current computational load to its aggregated computation capability:

\begin{equation}
    U_{i}(t) = \frac{T_{i,t}}{\phi_{i,t}}
    \label{eq:utilization}
\end{equation}

where $T_{i,t}$ represents the total computational load in the task queue of node $i$ at time $t$ measured in GFLOPs, and $\phi_{i,t}$ is the node's aggregated computation capability.

When deciding whether to transfer a task, node $i$ first identifies the neighboring node with minimum utilization:

\begin{equation}
    k^* = \underset{k \in M_i(t)}{\arg\min}\ U_{k}(t)
    \label{eq:minimum_neighbor}
\end{equation}

The task transfer decision is then governed by a piecewise function using the threshold parameter $\gamma$:

\begin{equation}
    \delta_{i \rightarrow k^*}(t) = 
    \begin{cases}
        \text{Transfer to node } k^*, & \text{if } U_i(t) - U_{k^*}(t) > \gamma \\
        \text{Process locally}, & \text{otherwise}
    \end{cases}
    \label{eq:delta_rule}
\end{equation}

The parameter $\gamma$ serves as a stability threshold that prevents excessive task transfers when utilization differences are minimal. A higher value of $\gamma$ makes the system more conservative, requiring a larger utilization gap to justify offloading, which reduces communication overhead but may lead to less balanced workloads. Conversely, a lower $\gamma$ makes the system more sensitive to utilization differences, promoting load balancing at the potential cost of increased communication overhead.

This mechanism ensures that tasks are only transferred when the utilization difference exceeds a meaningful threshold, preventing oscillatory behavior where tasks might be repeatedly transferred between nodes with similar utilization levels. Overall, this rule is the main idea of our distributed decision-making process, it provides simple but stable and efficient mechanism that avoids oscillatory offloading under dynamic swarm conditions.

\subsection{Congestion-Aware Early-Exit}

\label{sec:early_exit_mechanism}

To maintain consistent performance under varying workloads, nodes dynamically adjust the depth of the model inference in a congestion-aware manner based on real-time buffer conditions and the dynamics of the local task queue (the outstanding-task workload). This mechanism integrates a congestion-aware early-exit strategy that reduces computational load and latency by truncating the network execution at predefined layers when the outstanding workload increases, thereby producing intermediate outputs with acceptable accuracy levels while preventing excessive workload growth and latency spikes.

Figure~\ref{fig:arch} illustrates the proposed congestion-aware early-exit mechanism. Each node monitors the rate of change in its outstanding workload by computing the time-normalized derivative of the remaining task GFLOPs (a congestion indicator):

\begin{equation}
    \label{eq:derivative}
    \Delta T_{i,t} = \frac{T_{i,t} - T_{i,t-1}}{\Delta t},
\end{equation}
where \(\Delta T_{i,t}\) is the first difference of the remaining task GFLOPs and \(\Delta t\) is the interval between allocation steps. To account for abrupt fluctuations (e.g., due to new task creation or completion), we smooth this difference:

\begin{equation}
    \label{eq:derivative_smooth}
    D_{i,t} = D_{i,t} + \alpha(\Delta T_{i,t} - D_{i,t}),
\end{equation}

where \(D_{i,t}\) is the smoothed derivative and \(\alpha\) controls the smoothing factor. A high \(D_{i,t}\) value indicates sustained task accumulation, signaling that node \(v_i\) is unable to process its workload efficiently.

To prevent overload and reduce inference latency, each node adjusts its inference depth based on the smoothed derivative \(D_{i,t}\) of its task queue. Let \(\tau_{\mathrm{med}}\) and \(\tau_{\mathrm{high}}\) denote the medium and high congestion thresholds, respectively. The early-exit decision is characterized by the exit label \(\xi_{i,t}\), defined as:
\begin{equation}
    \xi_{i,t} =
    \begin{cases}
        L_{\text{full}}, & \text{if } D_{i,t} \le \tau_{\mathrm{med}}, \\
        L_1, & \text{if } \tau_{\mathrm{med}} < D_{i,t} \le \tau_{\mathrm{high}}, \\
        L_2, & \text{if } D_{i,t} > \tau_{\mathrm{high}},
    \end{cases}
    \label{eq:exit_rule}
\end{equation}
where \(L_{\text{full}}\) indicates processing the full network, \(L_1\) denotes truncation layer when there is medium congestion, and \(L_2\) is for the high congestion. In both early-exit cases (\(L_1\) and \(L_2\)), an additional three layers are processed after the exit point to finalize the output.

This adaptive mechanism ensures that nodes under high load degrade computation gracefully, trading off accuracy to maintain responsiveness and system stability. It is important to note that the neural network architecture, the selected split points, and the number of layers used in this work serve as a specific example corresponding to the evaluation presented in Section~\ref{sec:performance_evaluation}. In practice, the neural network designer must analyze the target model to identify appropriate split points and determine the number of additional layers to execute after each truncation. These decisions depend on the network's structure and the desired balance between accuracy and computational efficiency under dynamic conditions.

\color{black}
Algorithm~\ref{alg:dist_alloc_earlyexit} summarizes the proposed periodic decision logic at each node. It updates the aggregated computation capability, decides whether to process locally or offload to the least-utilized neighbor, and selects the inference depth via the congestion-aware early-exit rule. The per-epoch computational cost at node $v_i$ is dominated by neighbor-wise scans over $M_i(t)$ required by the $\arg\min$ utilization in Eq.~\ref{eq:minimum_neighbor} and the max term in Eq.~\ref{eq:agg_capability}, yielding $\mathcal{O}(|M_i(t)|)$ time; the early-exit update is $\mathcal{O}(1)$. The messaging overhead is $\mathcal{O}(|M_i(t)|)$ scalar exchanges with one-hop neighbors per epoch.
\color{black}

\begin{algorithm}[t]
\SetAlgoLined
\DontPrintSemicolon
{
\SetKwFunction{FMain}{DecisionEpoch}
\SetKwProg{Fn}{Function}{:}{}
\Fn{\FMain{$i,t$}}{

    \tcp{Update aggregated capability $\phi_{i,t+1}$ (Eq.~\ref{eq:agg_capability})}
    $\phi_{i,t+1} \leftarrow \textsc{UpdateAggCap}(i,t)$\;

    \tcp{Compute utilization $U_i(t)$ (Eq.~\ref{eq:utilization})}  
    $U_i \leftarrow \textsc{Utilization}(i,t)$\;
    \tcp{Find least utilized neighbor (Eq.~\ref{eq:minimum_neighbor})}  
    $k^* \leftarrow \arg\min\limits_{k\in M_i(t)} \textsc{Utilization}(k,t)$\;

    \tcp{Decide process vs.~offload (Eq.~\ref{eq:delta_rule})}
    $\delta \leftarrow \textsc{OffloadDecision}(U_i,\textsc{Utilization}(k^*,t),\gamma)$\;

    \If{$\delta=1$}{
        \textsc{Offload}$(\text{task},k^*)$\;
        \Return\;
    }

    \tcp{Update congestion indicator (Eqs.~\ref{eq:derivative}-\ref{eq:derivative_smooth})}
    $D_{i,t} \leftarrow \textsc{CongestionIndicator}(T_{i,t},T_{i,t-1},D_{i,t-1},\Delta t,\alpha)$\;

    \tcp{Select inference depth label $\xi_{i,t}$ (Eq.~\ref{eq:exit_rule})}
    $\xi_{i,t} \leftarrow \textsc{ExitLabel}(D_{i,t},\tau_{\mathrm{med}},\tau_{\mathrm{high}},L_{\text{full}},L_1,L_2)$\;

    \tcp{Continue execution from current state $l_i$}
    $\Delta t_{\mathrm{rem}} \leftarrow \Delta t$\;
    \While{$\Delta t_{\mathrm{rem}} > 0$ \textbf{and} $l_i \le \xi_{i,t}$}{
        $\Delta t_{\mathrm{rem}} \leftarrow \Delta t_{\mathrm{rem}} - \textsc{ComputeLayer}(l_i)$\;
        $l_i \leftarrow l_i+1$\;
    }
}
}
\caption{\color{black}Task allocation and congestion-aware early-exit at node $v_i$ (executed every $\Delta t$).}
\label{alg:dist_alloc_earlyexit}
\end{algorithm}

\begin{table}[htb]
    \centering
    \caption{Simulation Parameters}
    \begin{tabularx}{\columnwidth}{m{5.2cm}m{3cm}}
    \hline
        \textbf{Parameter}                      & \textbf{Value} \\ \hline    
        Number of Workers                       & 10-50 \\
        Area Bounds                             & 20x20 km \\
        Node Placement Granularity              & 15 \\
        Movement Radius                         & 1000 m \\
        Movement Speed                          & 75 m/s \\
        Worker Capability $(\mu, \sigma)$       & N(400,100) GFLOPs \\
        Energy per GFLOP                        & 0.02 J / GFLOP \\
        Task Generation Period                  & 60 ms \\
        Exit Points $(L_1, L_2, L_{\text{full}})$ & [15, 30, 60] \\
        Exit Thresholds $(\tau_{\mathrm{med}}, \tau_{\mathrm{high}})$                  & [1.5, 2.5] \\
        Exit Accuracy Levels                    & [0.6, 0.9, 0.95] \\
        Transmission Power                      & 30 dBm \\
        Noise Power                             & -85 dBm \\
        Minimum SNR ($\mathrm{SNR}_{\min}$)     & 3 dB \\
        Bandwidth                               & 10 MHz \\
        Number of Runs                          & 50 \\
        Maximum Simulation Time                 & 100 s \\
        Random Task Allocation Probability      & 0.2 \\
        Random Acyclic Offload Probability      & 0.1 \\
        Distributed Offloading $\gamma$         & 0.02 \\
        Decision Period                         & 200 ms \\
        Greedy Offloading Probability           & 0.05 \\
        \hline
    \end{tabularx}
    \label{tab:simParams}
\end{table}

\section{Performance Evaluation}
\label{sec:performance_evaluation}

Our simulation environment is designed to emulate the dynamic conditions of a large-scale UAV swarm executing split computing tasks. In our environment, UAVs are deployed over a 20×20 km area and follow circular trajectories with a movement radius of 1000 m, traveling at speeds of up to 75 m/s. Task arrivals are modeled as a Markov process with an average inter-arrival time of 60 ms, which accurately captures the random and bursty nature of workload generation. In addition, the simulator captures detailed communication and computation aspects,including realistic channel characteristics, transmission delays, and the heterogeneous computational capabilities of each UAV ensuring that both the physical movement and the fluctuating workload distribution among nodes are properly considered. Although the mission area is large, since the UAVs have high transmit power and predominantly have line-of-sight (LoS) links, and the network still remains largely connected for substantial portions of the flight. All the figures are presented with 95\% confidence intervals \color{black} derived from 50 independent runs. \color{black} Table~\ref{tab:simParams} shows the rest of the simulation parameters.

Within this dynamic simulation framework, we evaluate five distinct task offloading strategies. In the first scenario, tasks are offloaded in a purely random manner to a neighboring UAV, providing a basic baseline for performance comparison. The second strategy follows a similar random selection process, but enforces an acyclic constraint to prevent redundant offloading cycles. The third approach adopts a greedy policy, where each UAV offloads its task to the least busy neighbor based on instantaneous load measurements, thus aiming to minimize processing delays. The fourth strategy, referred to as LocalOnly, eliminates offloading entirely so that tasks are processed solely on the originating UAV, which highlights the trade-off between local processing overhead and the benefits of collaborative offloading. Finally, our proposed Distributed method leverages a diffusive metric, to dynamically route tasks based on both the local UAV’s capacity and the capabilities of its neighbors. This comprehensive set of cases allows us to rigorously assess how different offloading mechanisms perform under the challenging conditions of a mobile, Markov-driven UAV swarm.
\begin{figure}[hbt]
    \centering
    \includegraphics[width=0.38\textwidth]{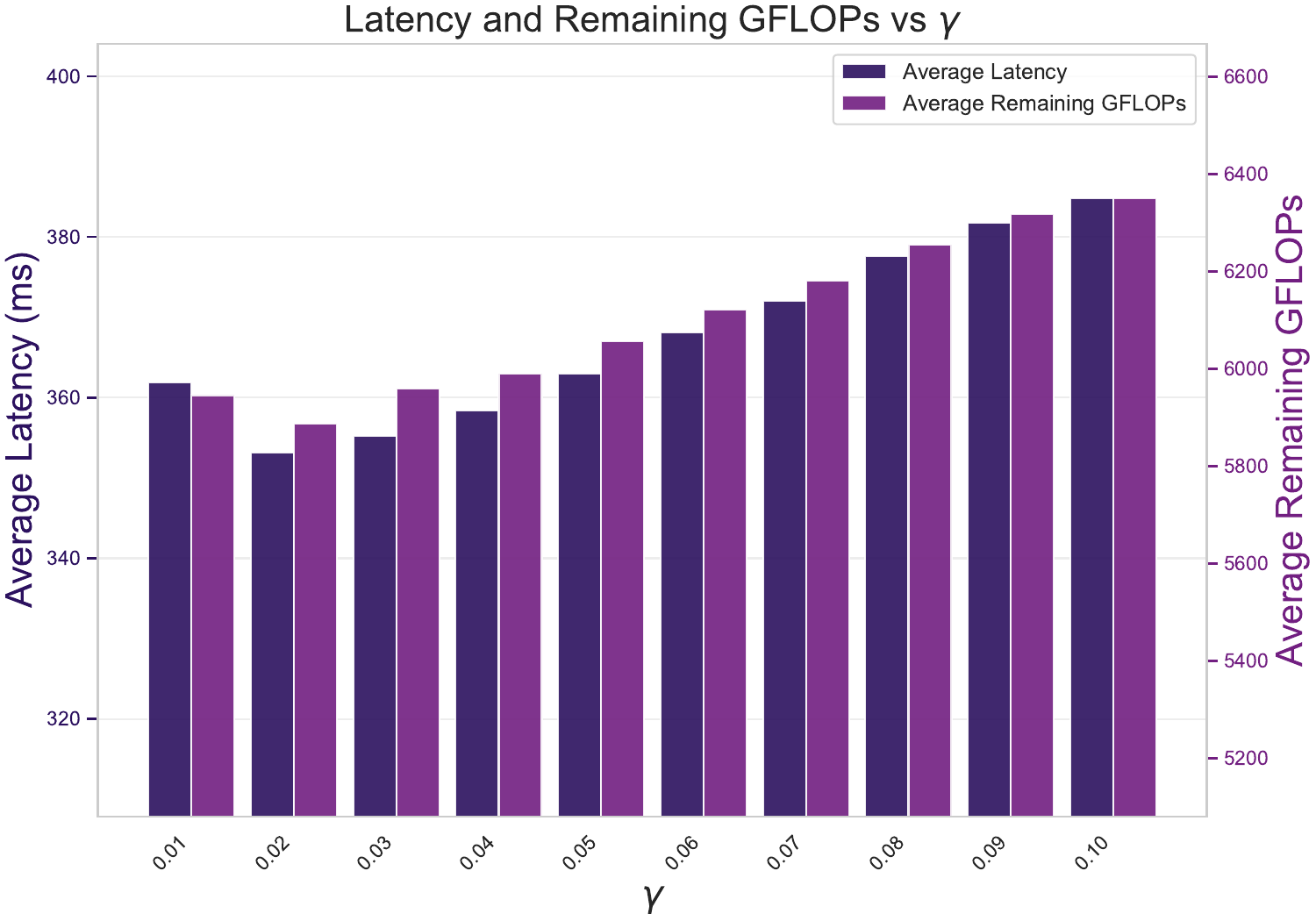}
    \caption{Sensitivity of the offloading threshold ($\gamma$) trade-off between average latency and outstanding workload (remaining GFLOPs).}
    \label{fig:gamma}
\end{figure}

In our experiments, we consider the neural network model described in Section \ref{sec:problem_formulation} pre-trained offline and deployed across the UAV swarm. The full model achieves an accuracy of 0.95. For congestion-aware depth control, we emulate two truncated inference operating modes (triggered by buffer dynamics in Section~4.2) using representative \emph{effective} accuracy levels of 0.9 (medium congestion) and 0.6 (high congestion). These values are simulation parameters intended to capture the expected quality/computation trade-off when inference is forced to terminate earlier. Whereas in practical early-exit systems, such operating points can be obtained via validation-set calibration using confidence/uncertainty thresholds~\cite{khademsohiselfxit} or considered under anytime/forced-exit settings~\cite{yang2023adadet,kuhse2025you}. Lastly, we set the distributed allocation threshold to $\gamma=0.02$ based on the sensitivity analysis in figure~\ref{fig:gamma}. We selected this value to effectively balance the trade-off between frequent task transfers and achieving equitable workload distribution.

\begin{figure*}[htb]
    \centering
    \begin{subfigure}[b]{0.32\textwidth}
        \includegraphics[width=1.1\linewidth]{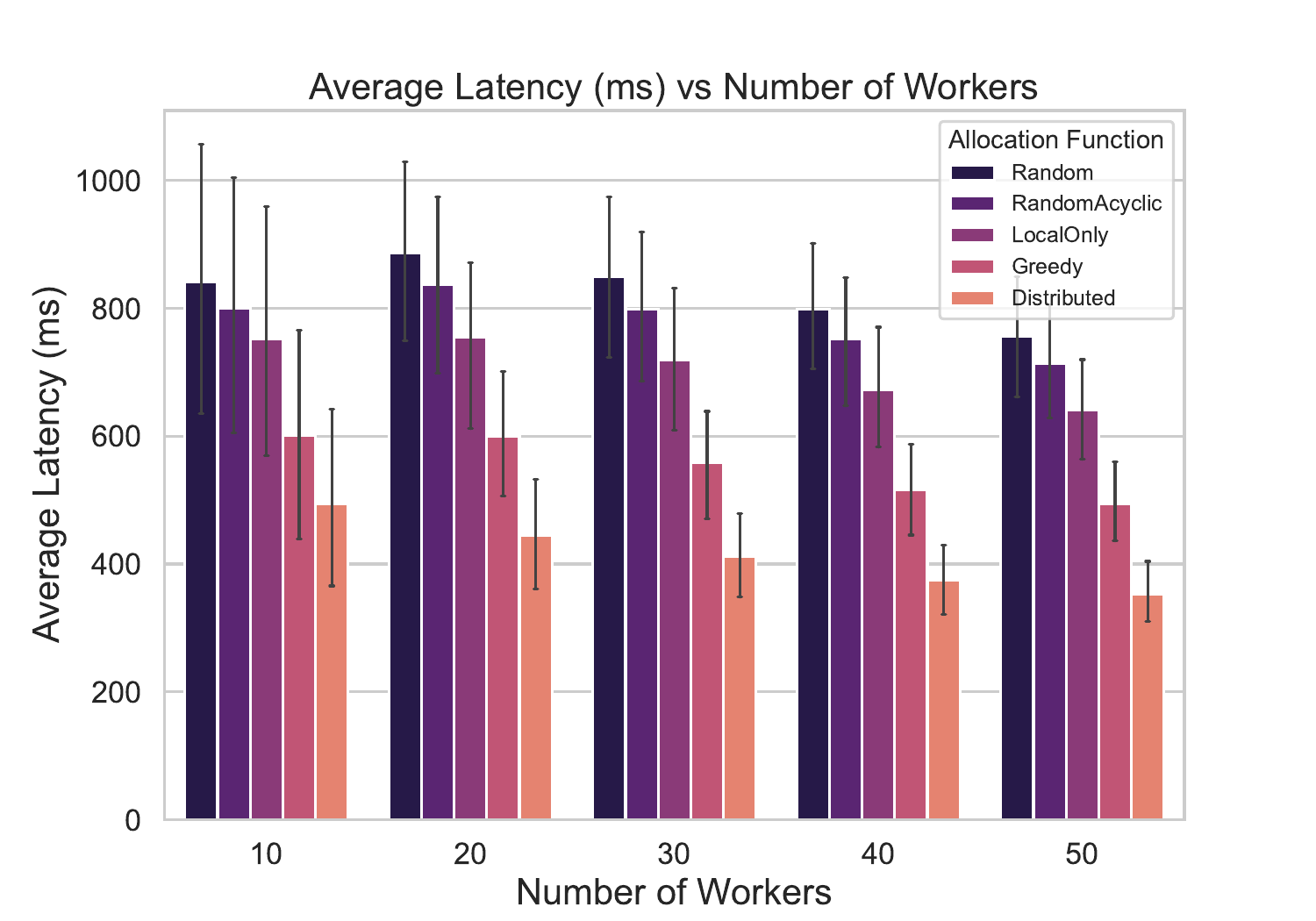}
        \caption{}
        \label{fig:basic_latency}
    \end{subfigure}
    \hfill
    \begin{subfigure}[b]{0.32\textwidth}
        \includegraphics[width=1.1\linewidth]{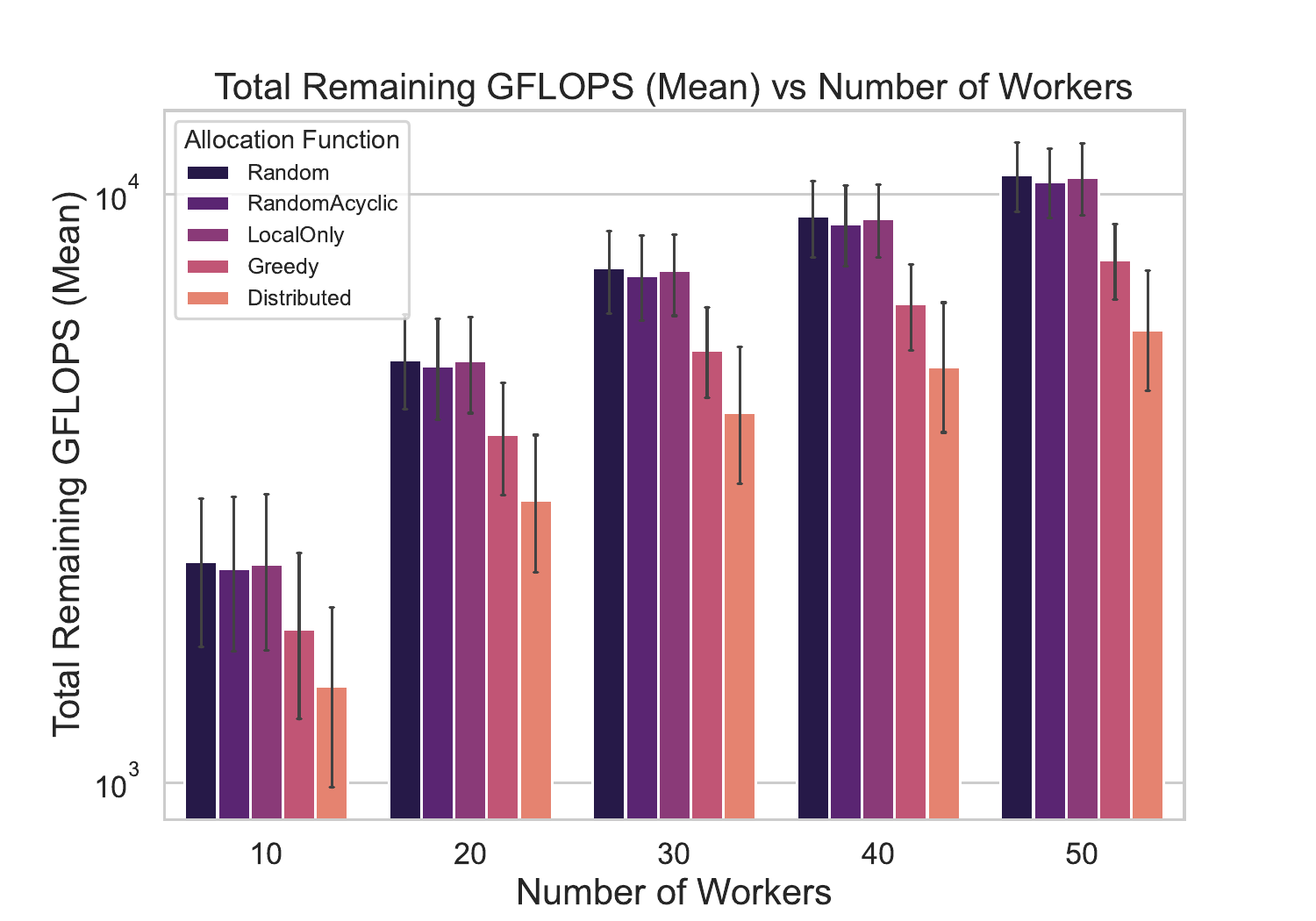}
        \caption{}
        \label{fig:basic_gflops}
    \end{subfigure}
    \hfill
    \begin{subfigure}[b]{0.32\textwidth}
        \includegraphics[width=1.1\linewidth]{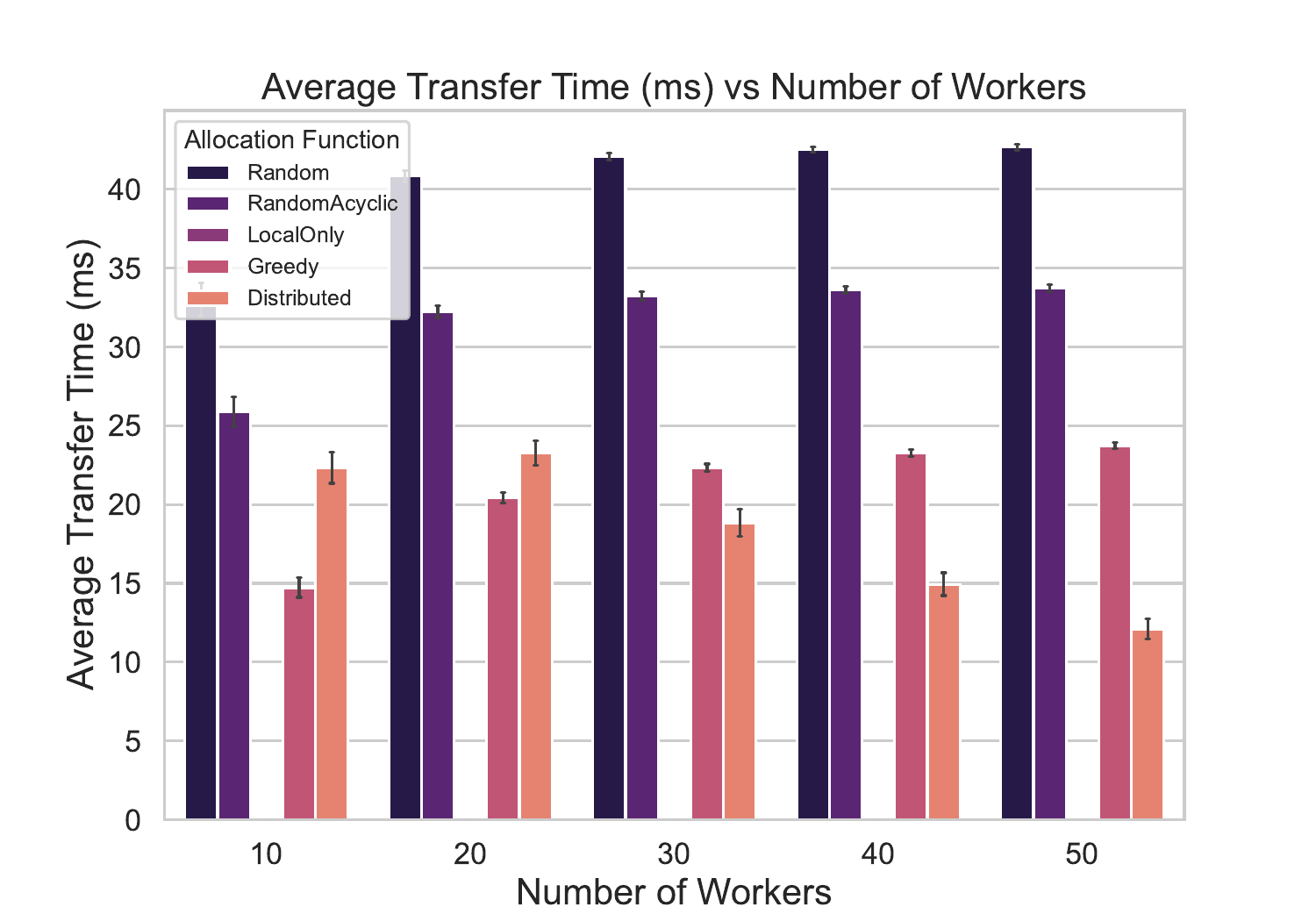}
        \caption{}
        \label{fig:basic_transfer}
    \end{subfigure}
    \vskip\baselineskip
    \begin{subfigure}[b]{0.32\textwidth}
        \includegraphics[width=1.1\linewidth]{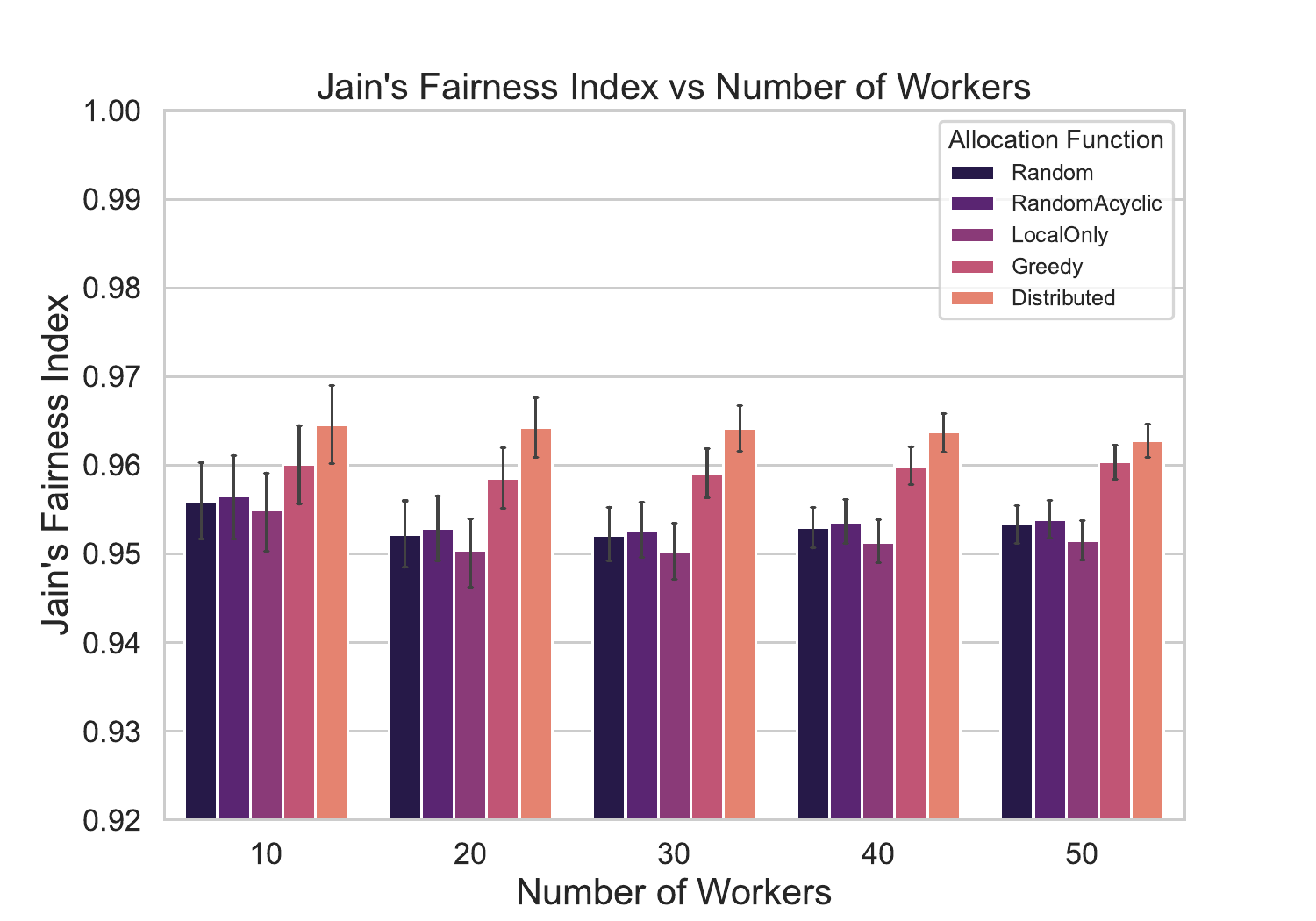}
        \caption{}
        \label{fig:basic_fairness}
    \end{subfigure}
    \hfill
    \begin{subfigure}[b]{0.32\textwidth}
        \includegraphics[width=1.1\linewidth]{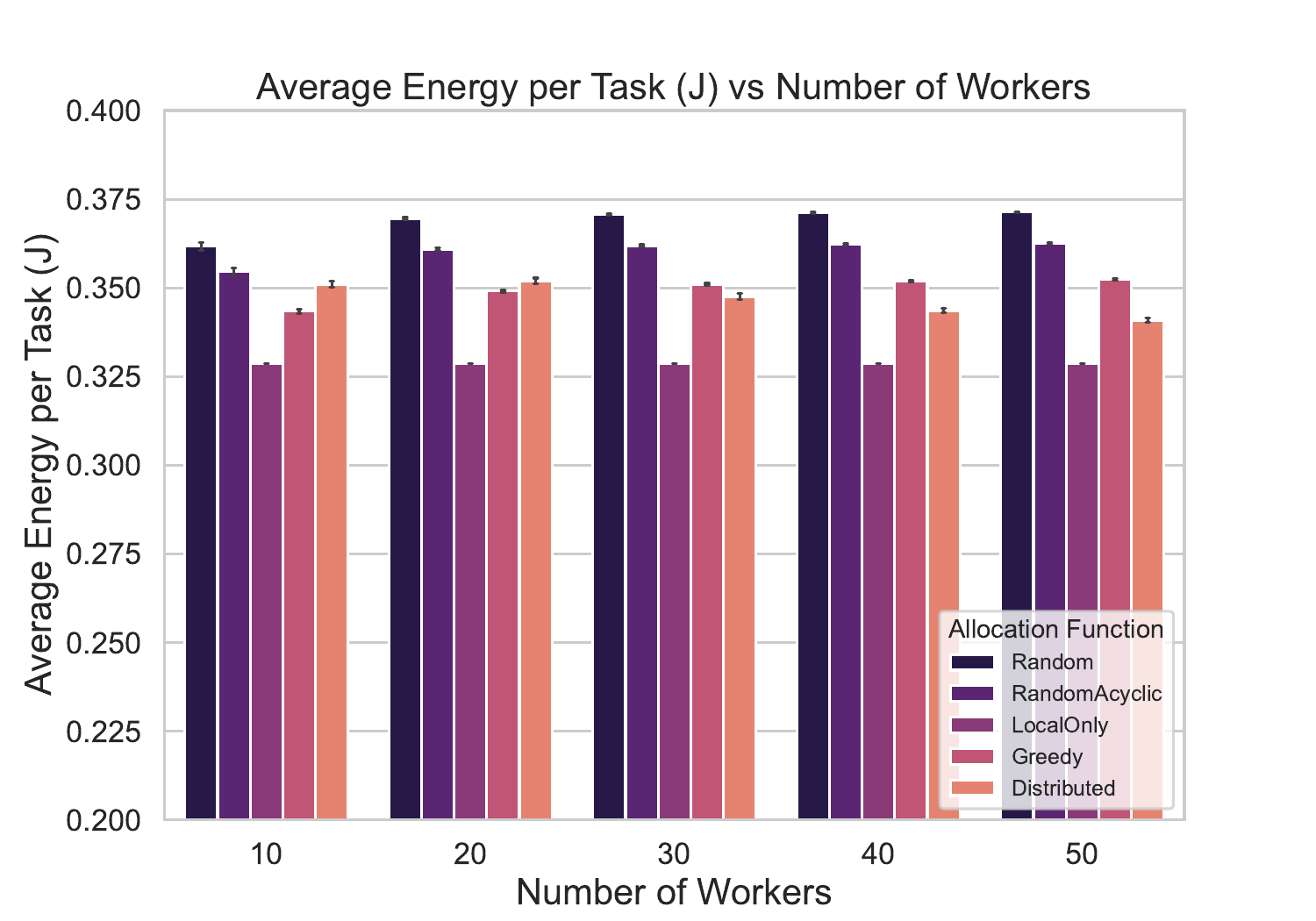}
        \caption{}
        \label{fig:basic_energy}
    \end{subfigure}
    \hfill
    \begin{subfigure}[b]{0.32\textwidth}
        \includegraphics[width=1.1\linewidth]{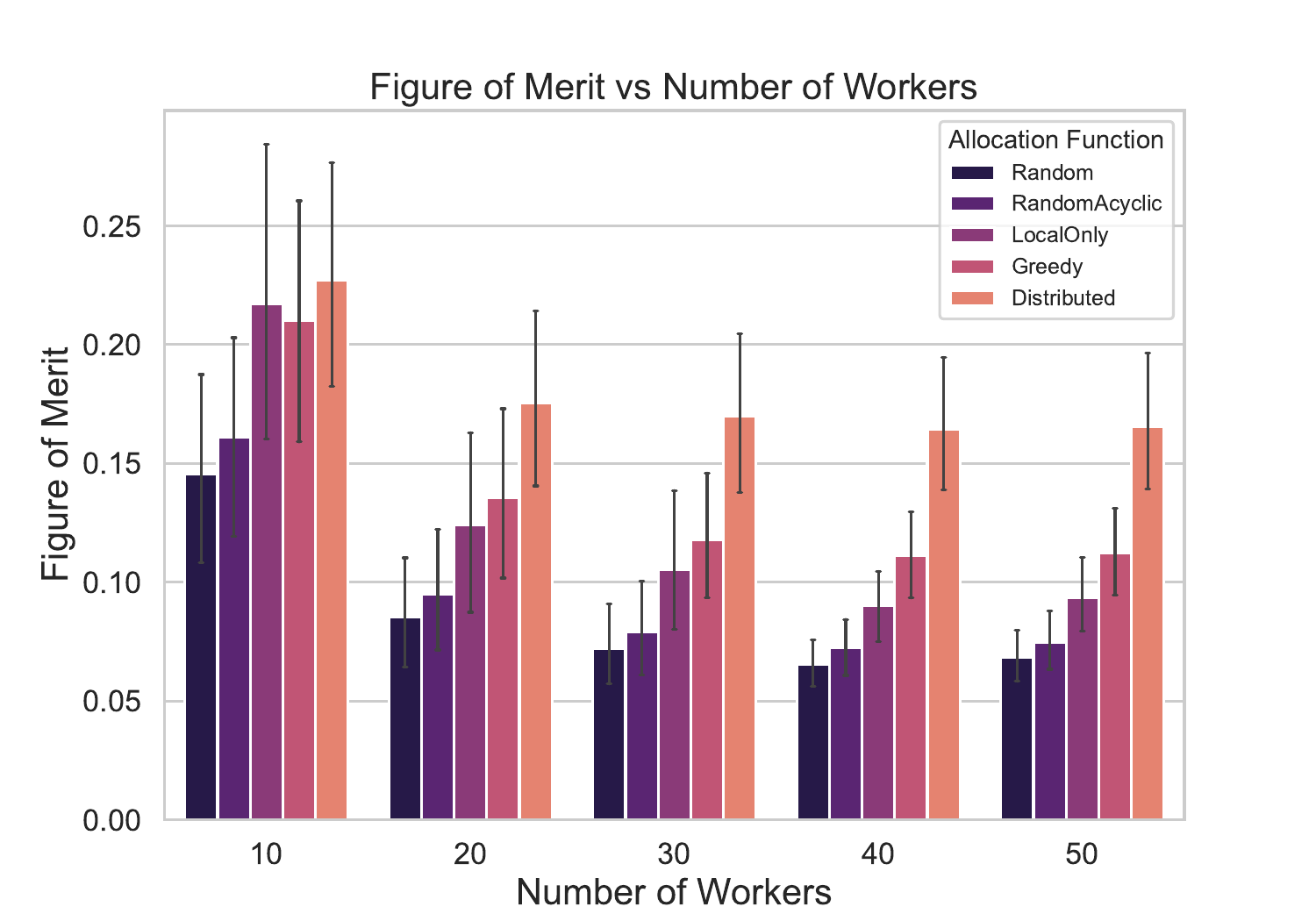}
        \caption{}
        \label{fig:basic_merit}
    \end{subfigure}
    \caption{Simulation results as a function of the number of workers. (a) shows average latency, (b) depicts remaining GFLOPs, (c) illustrates average transfer time, (d) presents Jain's fairness index, (e) reports energy per task, and (f) shows the overall figure of merit. \color{black} All figures are presented with 95\% confidence intervals.}
    \label{fig:basic_group}
\end{figure*}

\subsection{Basic Simulations}

The initial set of simulations examines a baseline scenario that does not employ early-exit mechanism. Figure~\ref{fig:basic_latency} illustrates the average task latency as a function of the number of workers, demonstrating that latency decreases as more workers are available. In Figure~\ref{fig:basic_gflops}, the average remaining GFLOPs in UAV buffers are depicted, providing insight into workload accumulation and balancing among nodes. Figure~\ref{fig:basic_transfer} presents the average transfer times, where the greedy approach minimizes transfers initially but experiences an increase as load builds, while the distributed method transfers only when necessary. Jain's fairness index is reported in Figure~\ref{fig:basic_fairness} by normalizing the processed GFLOPs with respect to each UAV's capability, thereby ensuring that more capable nodes receive a proportionally higher workload. Figure~\ref{fig:basic_energy} shows the energy consumption per task, where the local-only method benefits from avoiding transmission overhead.

Lastly, we come up with a figure of merit (FOM) that combines previously measured metrics to gauge the overall performance of the methods in (\ref{eq:merit}). The FOM is measured in Figure~\ref{fig:basic_merit} and illustrates that our proposed distributed method consistently achieves the highest performance across all worker counts, although the local-only approach remains competitive at lower worker densities due to a reduced number of generated tasks.
\begin{equation}
    \label{eq:merit}
    \text{FOM} = \frac{\text{TPS} \times \text{ACC} }{\text{AE} \times \text{AL}}
\end{equation}
where TPS is task completed per second, ACC is average accuracy of completed tasks, AE is average energy usage, and AL is average latency per task.

Overall, the results presented in Figures~\ref{fig:basic_group} collectively demonstrate the effectiveness of the proposed distributed method in balancing key performance trade-offs across varying swarm sizes. Specifically, the distributed approach achieves consistently lower latency, reduced workload accumulation, and minimal transfer overhead compared to alternative strategies. In addition, it exhibits a narrower latency confidence interval, indicating lower jitter and more stable responsiveness under dynamic conditions. Moreover, it maintains a high degree of fairness in task distribution and favorable energy consumption, despite the added complexity of inter-node coordination. These advantages become more pronounced as the number of workers increases, highlighting the scalability of the proposed solution. The superiority of the distributed method is further confirmed by the FOM index, which integrates all core performance metrics into a single indicator and consistently favors the proposed scheme across the entire range of worker densities. These findings underscore the capability of our system to efficiently manage computational loads in dynamic UAV environments without centralized control, while ensuring both high resource utilization and robustness against congestion.

\begin{figure*}[htb]
    \centering
    \begin{subfigure}[b]{0.32\textwidth}
        \includegraphics[width=1.1\linewidth]{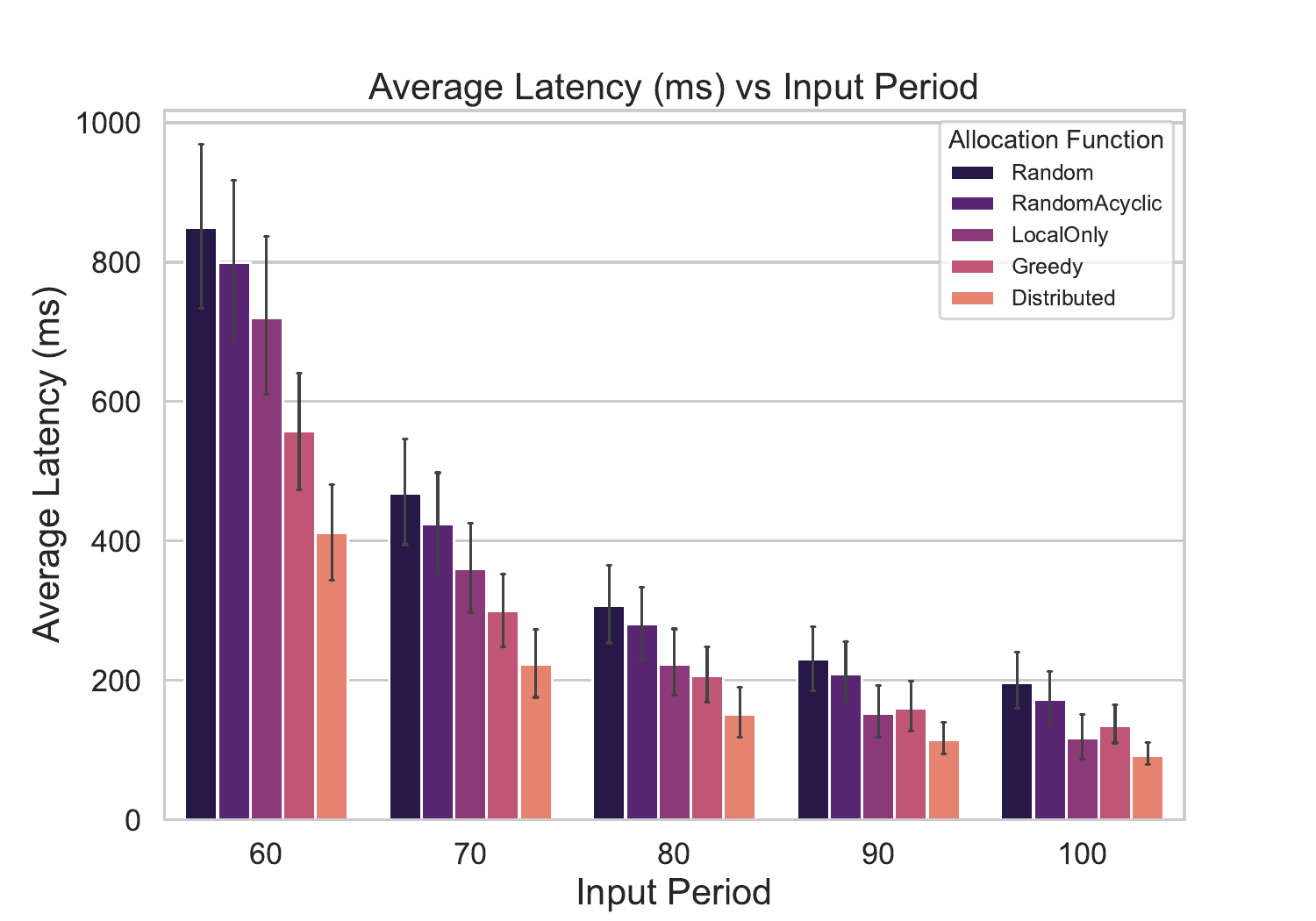}
        \caption{}
        \label{fig:input_latency}
    \end{subfigure}
    \hfill
    \begin{subfigure}[b]{0.32\textwidth}
        \includegraphics[width=1.1\linewidth]{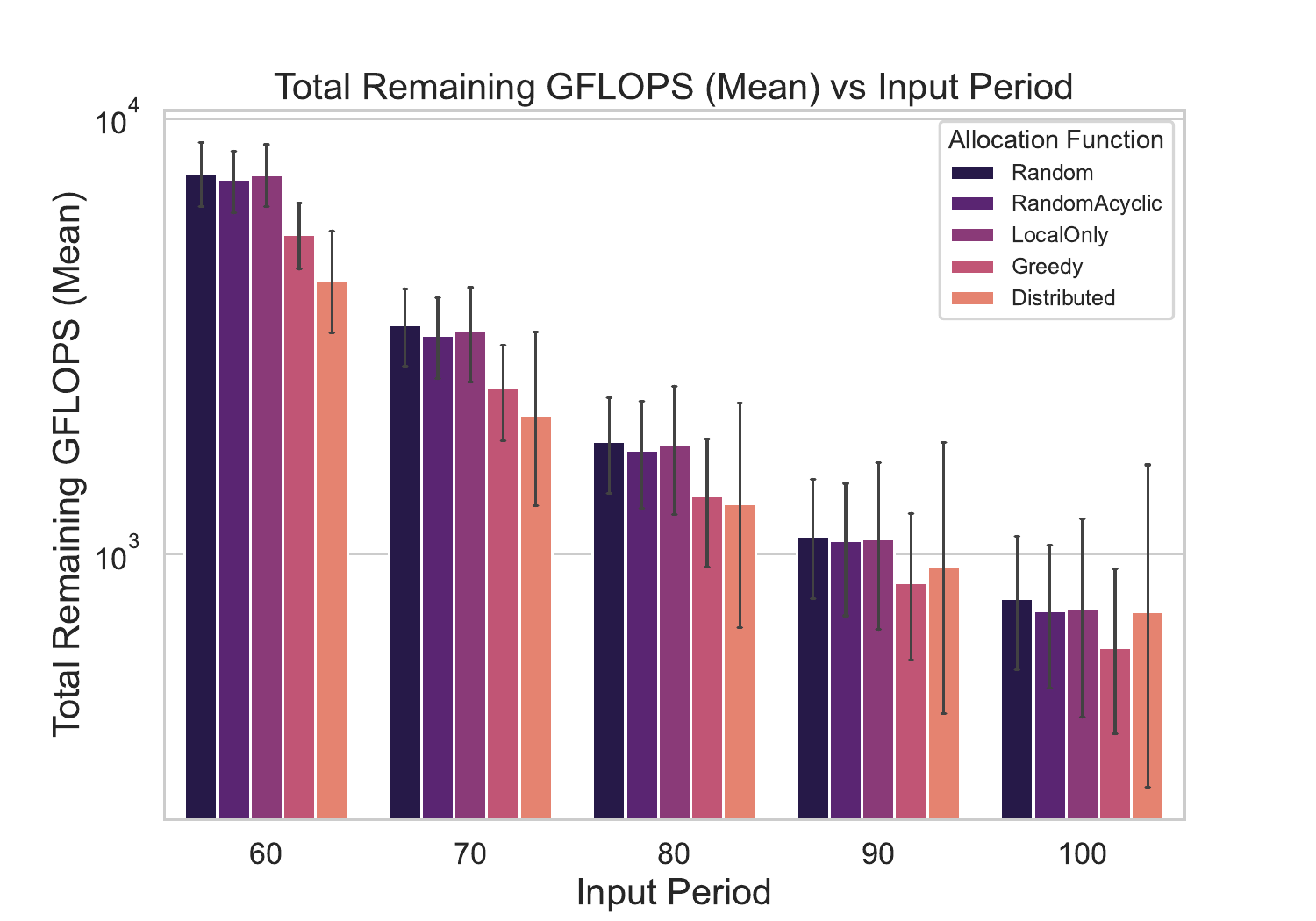}
        \caption{}
        \label{fig:input_gflops}
    \end{subfigure}
    \hfill
    \begin{subfigure}[b]{0.32\textwidth}
        \includegraphics[width=1.1\linewidth]{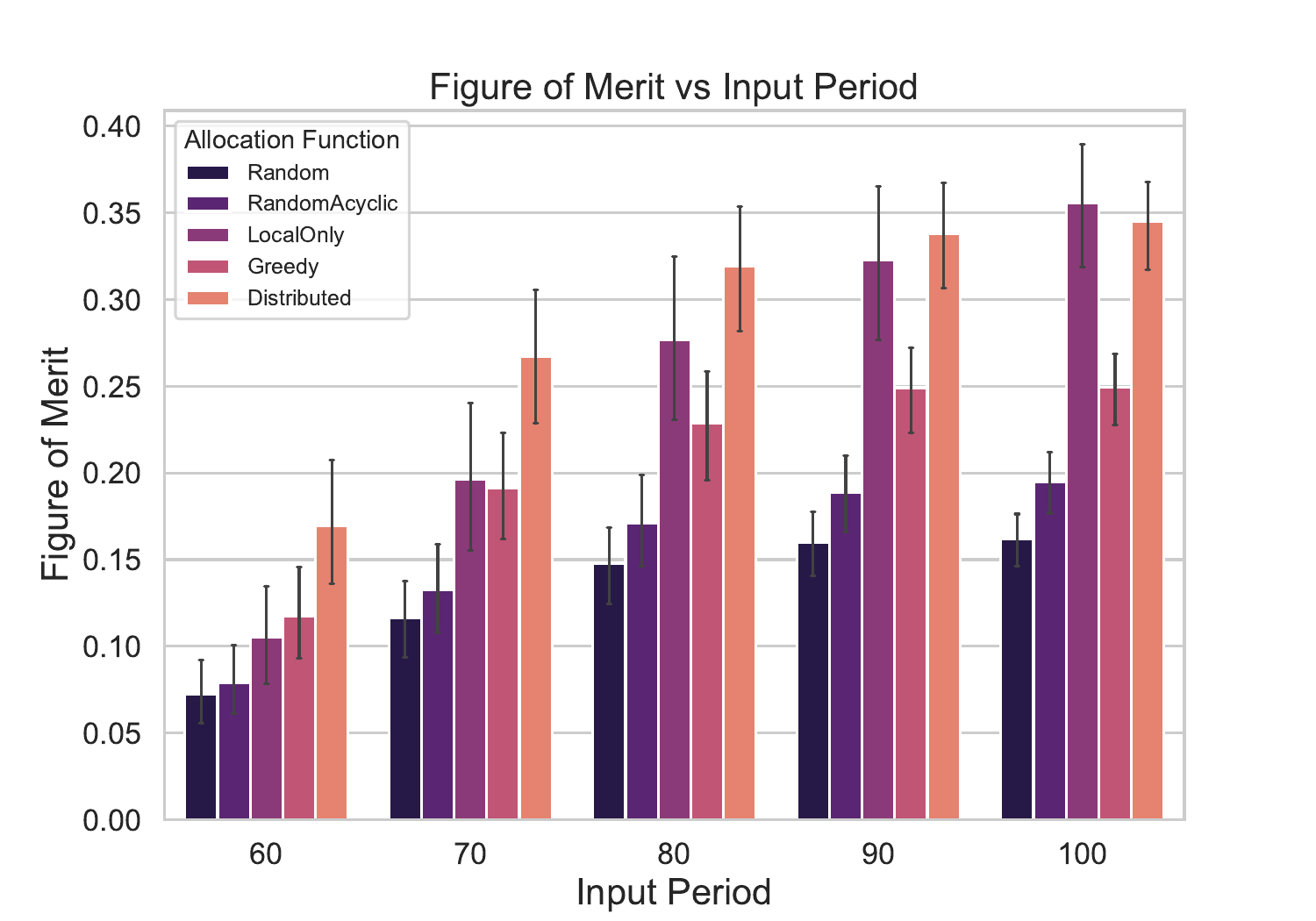}
        \caption{}
        \label{fig:input_merit}
    \end{subfigure}
    \caption{Performance metrics under varying task arrival rates for 30 workers. Figure (a) shows average latency, (b) displays remaining GFLOPs, and (c) presents the overall figure of merit. \color{black} All figures are presented with 95\% confidence intervals.}
    \label{fig:input_group}
\end{figure*}

\begin{figure*}[htb]
    \centering
    \begin{subfigure}[b]{0.32\textwidth}
        \includegraphics[width=1.1\linewidth]{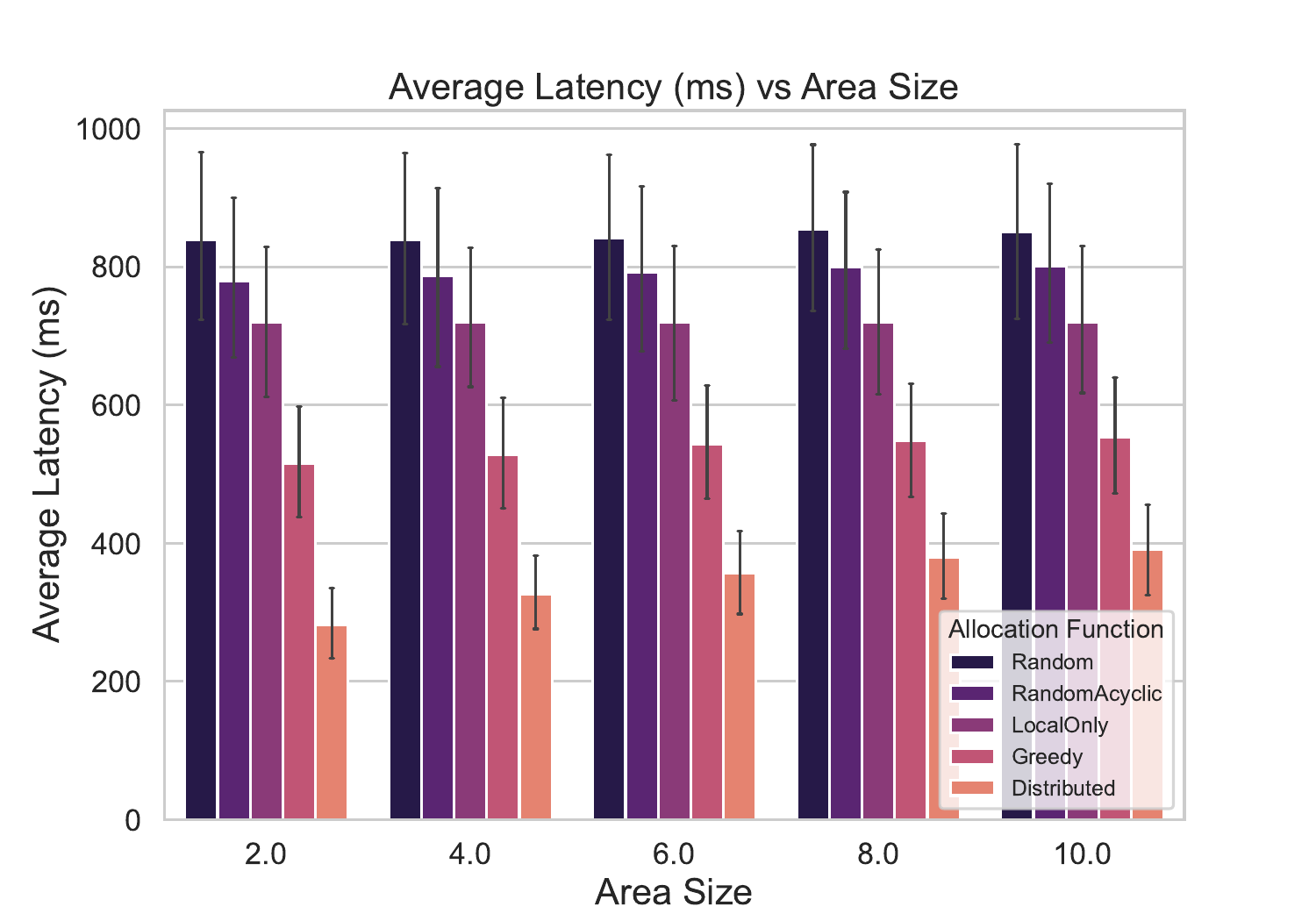}
        \caption{}
        \label{fig:area_latency}
    \end{subfigure}
    \hfill
    \begin{subfigure}[b]{0.32\textwidth}
        \includegraphics[width=1.1\linewidth]{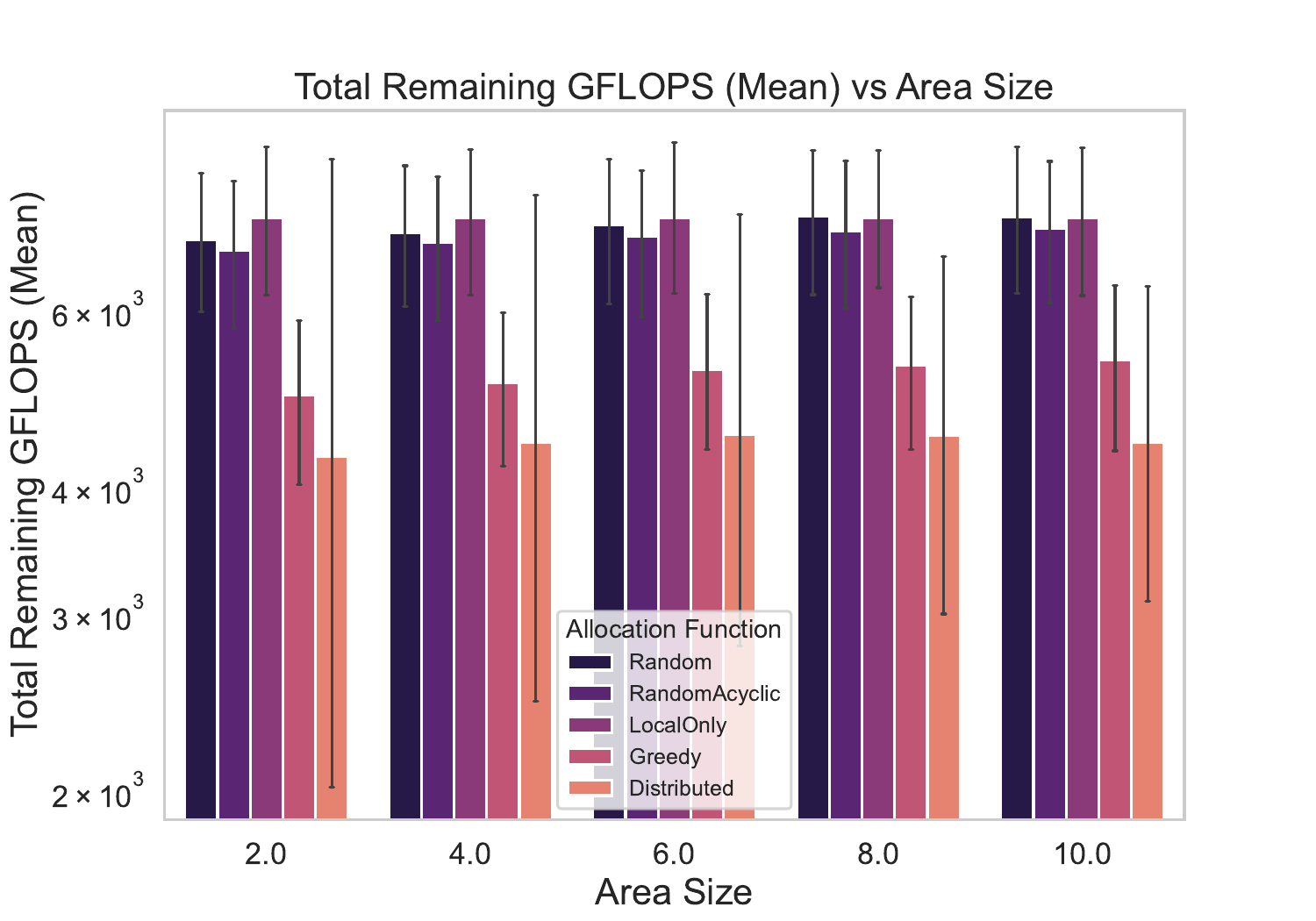}
        \caption{}
        \label{fig:area_gflops}
    \end{subfigure}
    \hfill
    \begin{subfigure}[b]{0.32\textwidth}
        \includegraphics[width=1.1\linewidth]{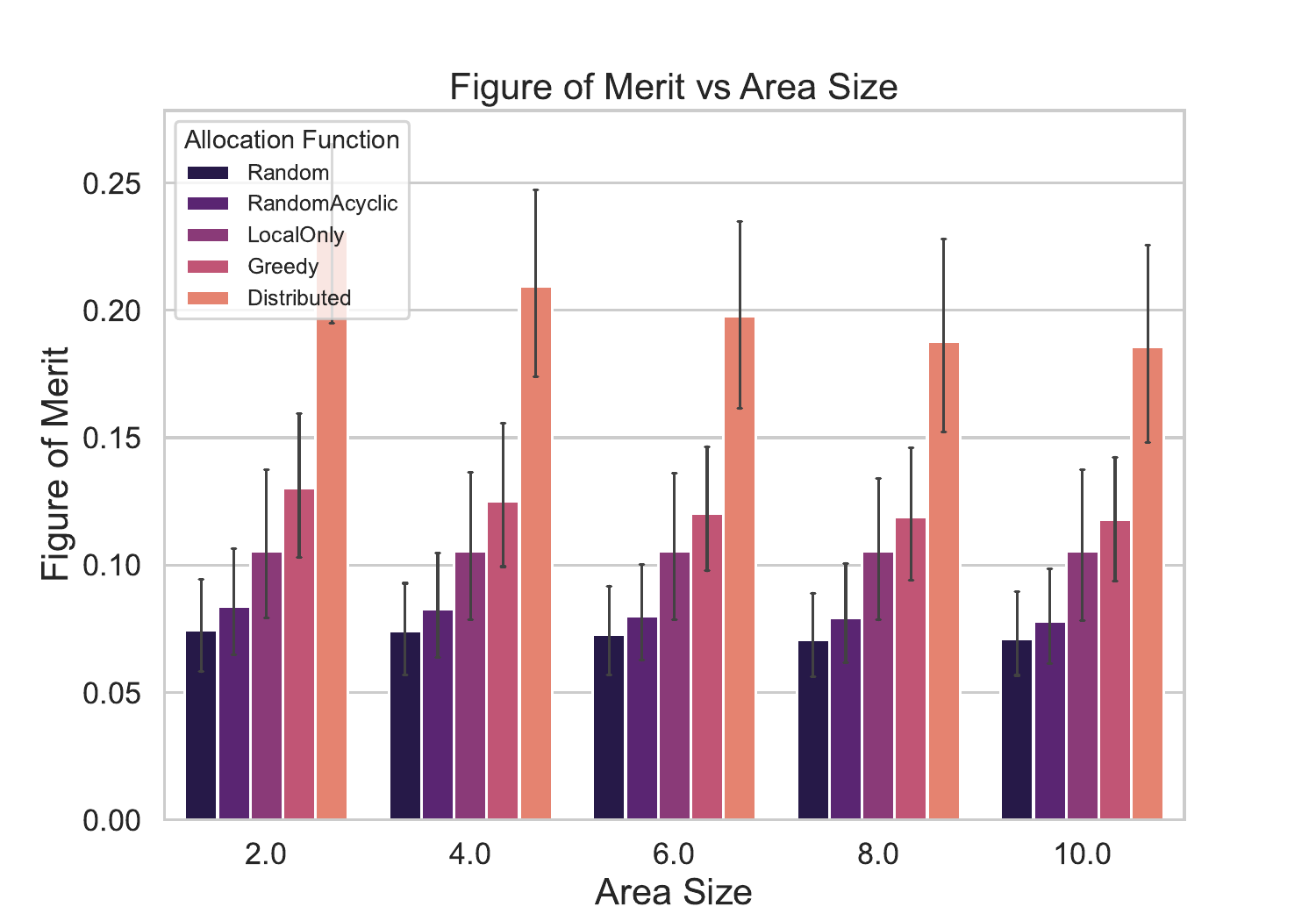}
        \caption{}
        \label{fig:area_merit}
    \end{subfigure}
    \caption{Performance metrics under varying mission area sizes. Figure (a) shows average latency, (b) depicts remaining GFLOPs, and (c) presents the overall figure of merit. \color{black} All figures are presented with 95\% confidence intervals.}
    \label{fig:area_group}
\end{figure*}

\subsection{Effect of Input Rate}

Next, we measure the effect of input rate when there is fixed number of workers. We vary mean task arrival time from 60ms to 100ms for 30 workers. Figure \ref{fig:input_latency} has the average task latency results. Our proposed method outperforms others for high task input rates (lower input periods). However as task arrival time increases, the gap between LocalOnly and out method diminishes. Because, at this point most of the edge nodes are idle and are not overloaded. Thus, overhead of transmitting tasks actually hurt the performance as in Random methods. As a result, our method cannot find opportunities to collaborate, which makes it perform similar to LocalOnly.
Moreover, in figure \ref{fig:input_gflops} we take a look into remaining task GFLOPs in edge node buffers. The behavior is similar to latency results. One thing to notice here is that lower latency does not result in lower remaining task GFLOPs. Because a task can be half processed, have less remaining GFLOPs and still completed later. The difference of LocalOnly and Random methods illustrate this clearly.
Lastly, we measure the FOM for all methods in figure \ref{fig:input_merit}. When the network is stressed, the LocalOnly method cannot keep up with Greedy and our Distributed method. One interesting result here is that, when the network is relaxed the LocalOnly is the overall best method. This makes sense because when there are not many tasks, the overhead of collaboration does provide significant benefits.

\subsection{Effect of Network Distribution}

The performance of collaborative task offloading is further influenced by the spatial distribution of UAVs, as captured by variations in the mission area size. Figure~\ref{fig:area_latency} shows that as the mission area increases, the opportunities for inter-node collaboration decline, resulting in increased overall latency. This effect is particularly pronounced in larger areas where the network connectivity weakens, causing the system to behave more like a collection of isolated local processors rather than an integrated distributed system. The impact on workload balancing is further illustrated in Figure~\ref{fig:area_gflops}, where larger mission areas correlate with a higher average of remaining GFLOPs in certain nodes, indicating that the collaborative mechanism is less effective in redistributing the load uniformly. Consequently, the overall figure of merit (Figure~\ref{fig:area_merit}) trends downward as the mission area expands, underscoring the sensitivity of the distributed strategy to network density and connectivity. These observations suggest that for large-scale deployments, additional strategies to enhance connectivity or adaptive routing protocols may be necessary to sustain optimal performance.

\begin{figure*}[htb]
    \centering
    \begin{subfigure}[b]{0.32\textwidth}
        \includegraphics[width=1.1\linewidth]{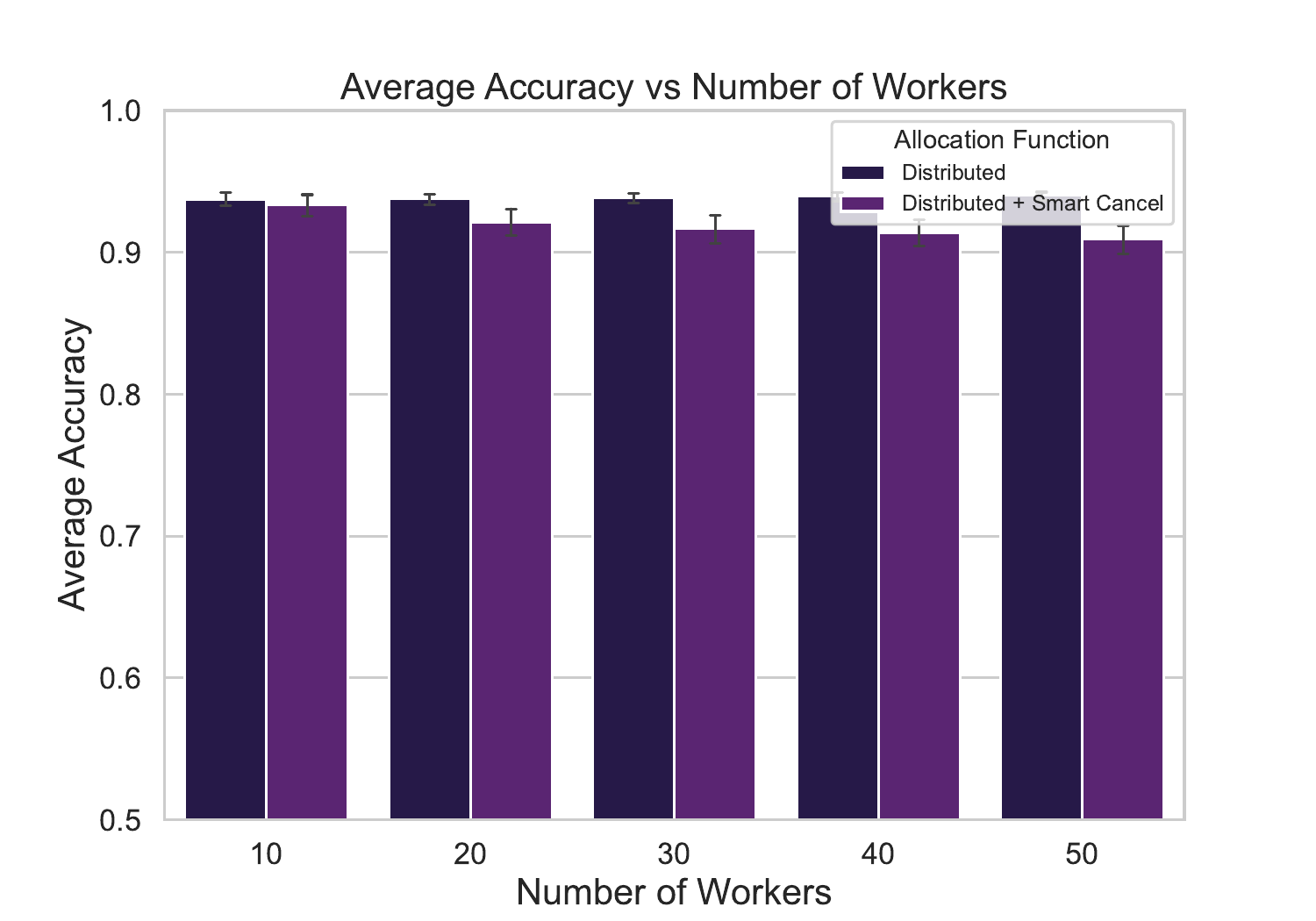}
        \caption{}
        \label{fig:dynamic_accuracy}
    \end{subfigure}
    \hfill
    \begin{subfigure}[b]{0.32\textwidth}
        \includegraphics[width=1.1\linewidth]{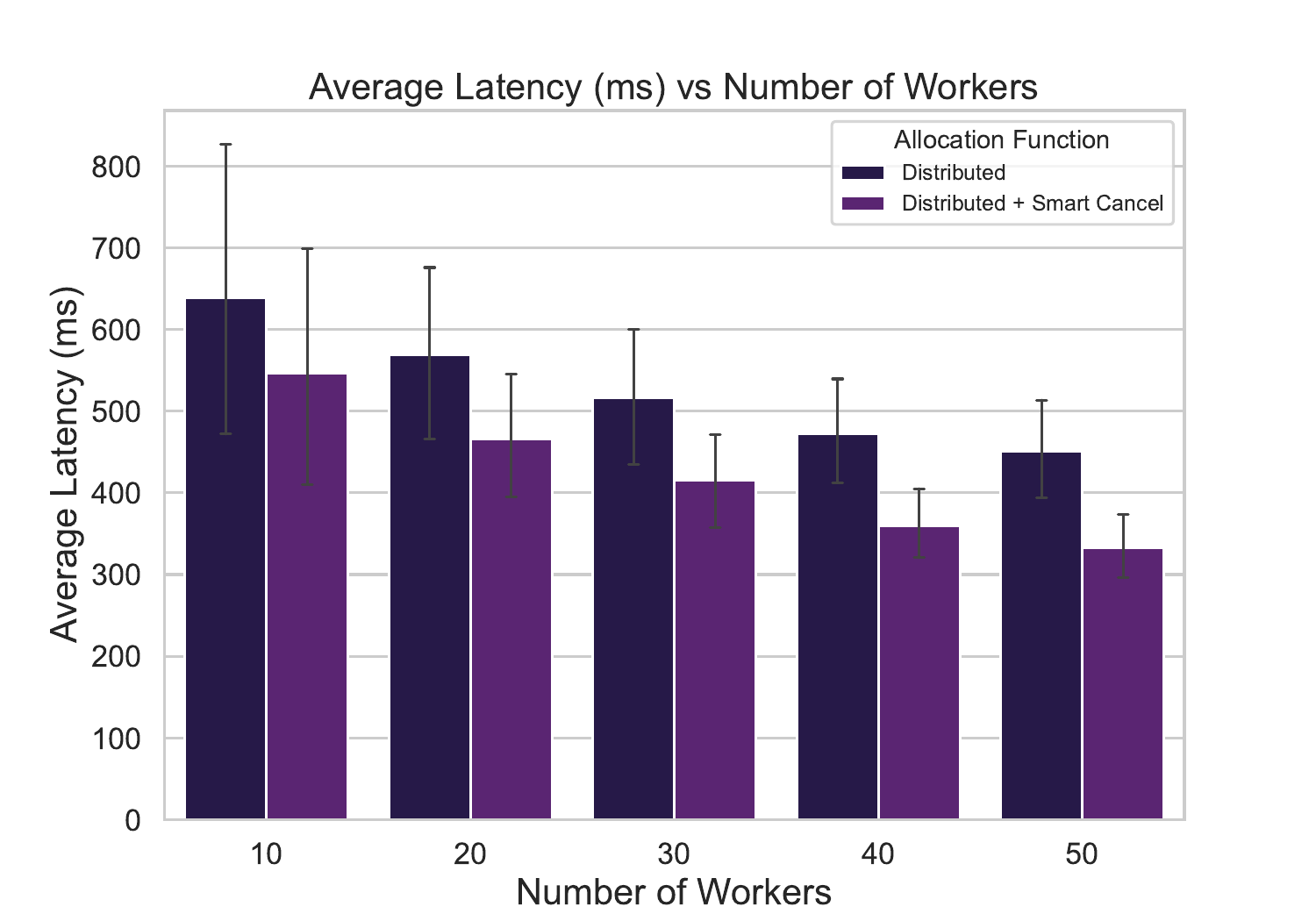}
        \caption{}
        \label{fig:dynamic_latency}
    \end{subfigure}
    \hfill
    \begin{subfigure}[b]{0.32\textwidth}
        \includegraphics[width=1.1\linewidth]{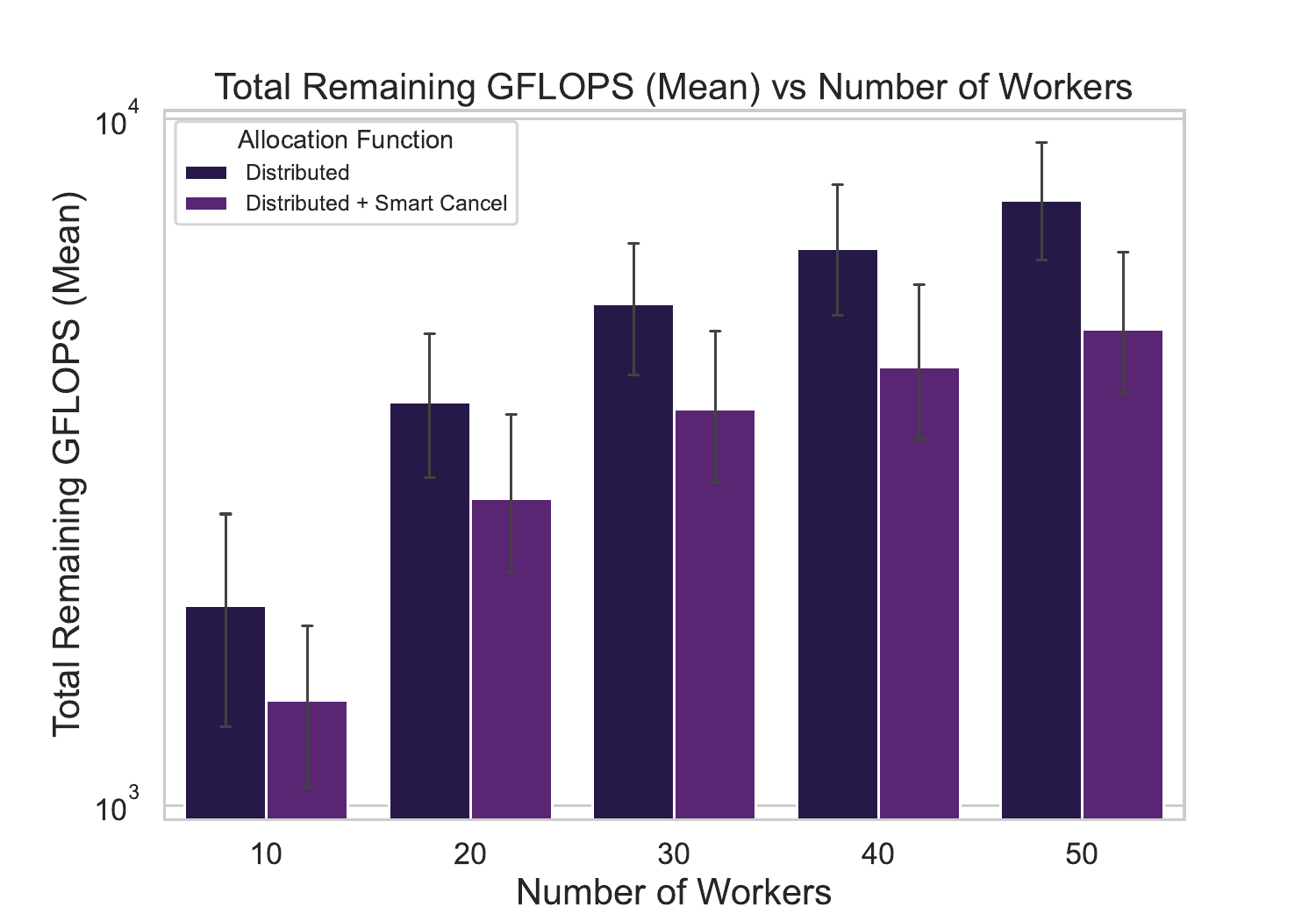}
        \caption{}
        \label{fig:dynamic_gflops}
    \end{subfigure}
    \vskip\baselineskip
    \begin{subfigure}[b]{0.32\textwidth}
        \includegraphics[width=1.1\linewidth]{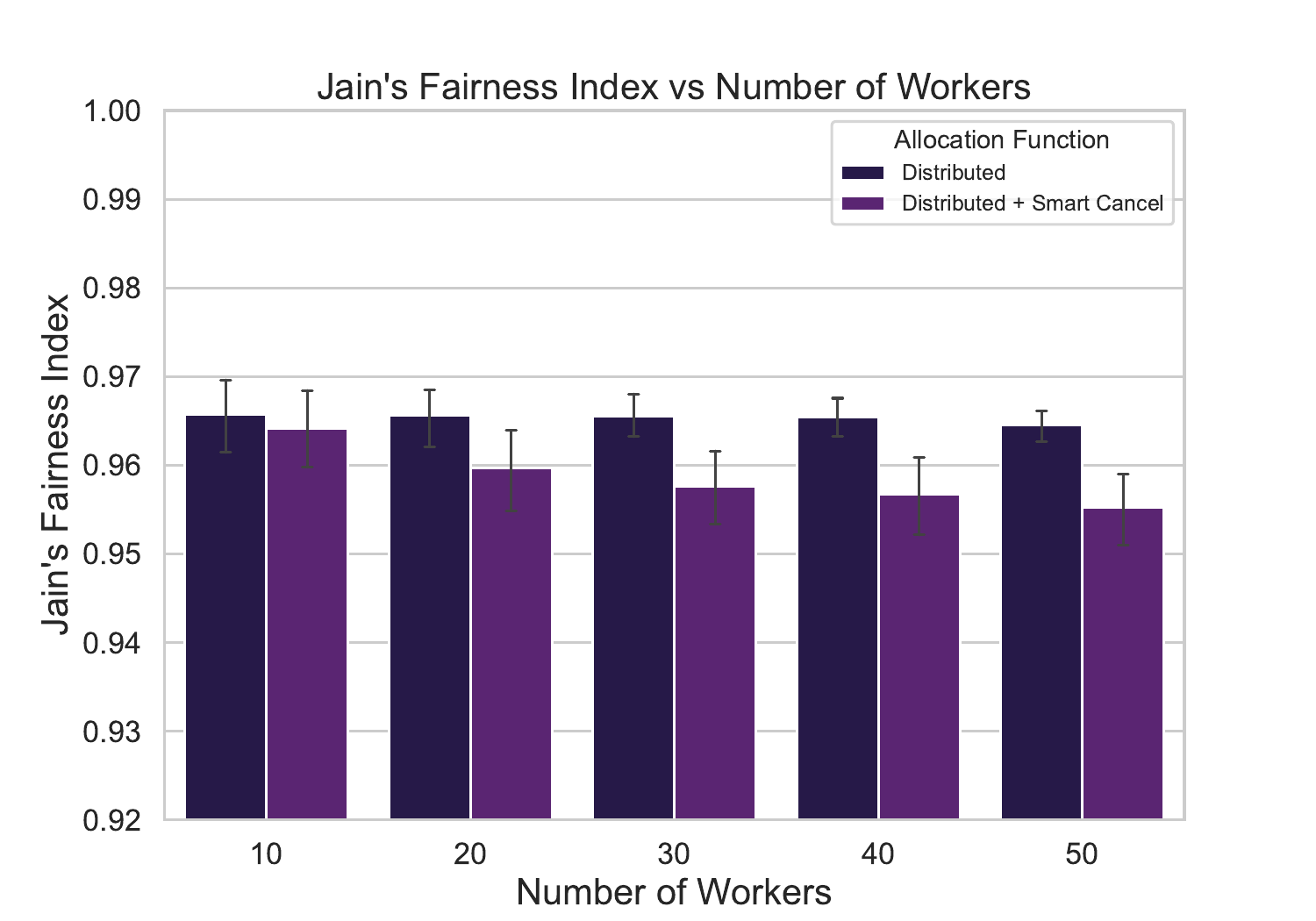}
        \caption{}
        \label{fig:dynamic_fairness}
    \end{subfigure}
    \hfill
    \begin{subfigure}[b]{0.32\textwidth}
        \includegraphics[width=1.1\linewidth]{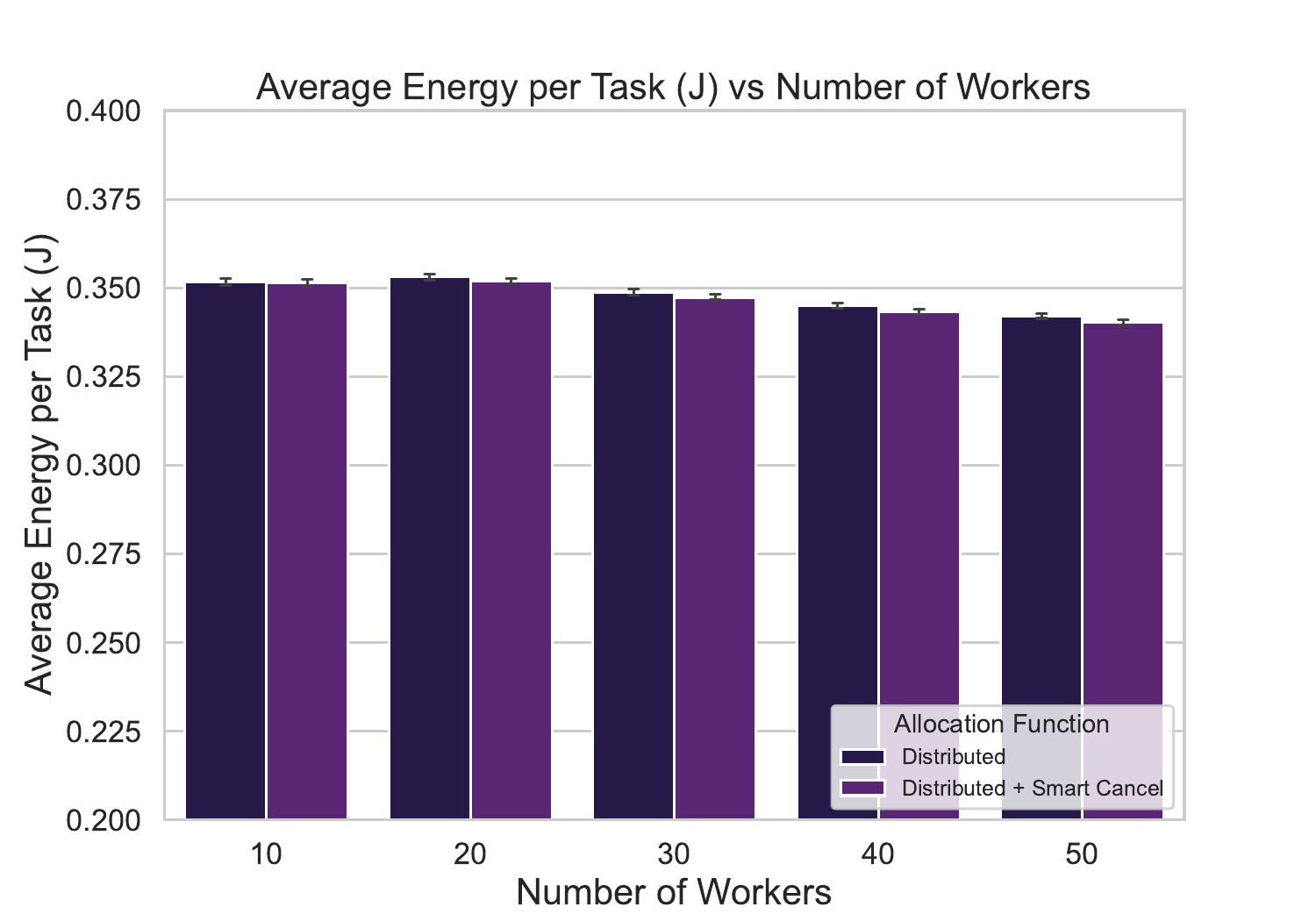}
        \caption{}
        \label{fig:dynamic_energy}
    \end{subfigure}
    \hfill
    \begin{subfigure}[b]{0.32\textwidth}
        \includegraphics[width=1.1\linewidth]{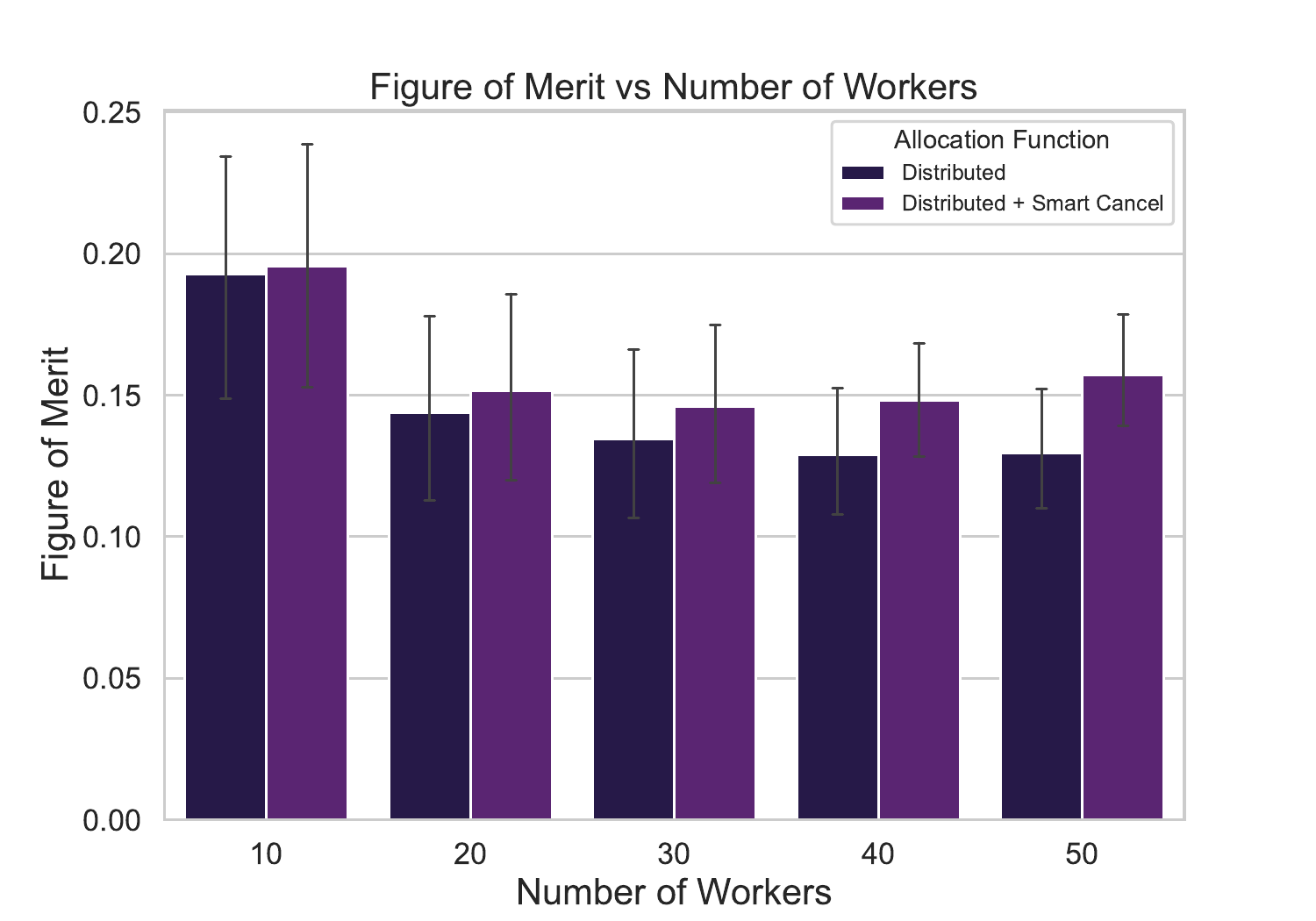}
        \caption{}
        \label{fig:dynamic_merit}
    \end{subfigure}
    \caption{congestion-aware Early-exit mechanism performance as a function of the number of workers. Figure (a) shows average accuracy, (b) depicts average latency, (c) illustrates remaining GFLOPs, (d) shows the fairness index, (e) presents energy per task, and (f) shows the overall figure of merit. \color{black} All figures are presented with 95\% confidence intervals.}
    \label{fig:dynamic_group}
\end{figure*}

\subsection{Congestion-aware Early-Exit Mechanism}

In the final set of experiments, we evaluate the efficacy of our congestion-aware early-exit mechanism, which adapts the network configuration in real-time based on load and buffer conditions executing an early-exit mechanism. This mechanism is designed to preemptively mitigate bottlenecks by allowing overloaded nodes to reduce their local computation and offload tasks to less burdened neighbors. The benefits of this approach are clearly evident across multiple metrics. For instance, Figure~\ref{fig:dynamic_accuracy} demonstrates that congestion-aware early-exit helps maintain high inference accuracy by preventing nodes from becoming overwhelmed. Similarly, the average latency, as shown in Figure~\ref{fig:dynamic_latency}, is significantly reduced when early-exit is enabled. The reduction in latency can be attributed to the more efficient distribution of computational tasks, which minimizes the time that tasks spend waiting in overloaded buffers.

Figure~\ref{fig:dynamic_gflops} provides insights into the distribution of remaining GFLOPs across the network, highlighting thatcongestion-aware early-exit results in a more even distribution of workload. This balance is further corroborated by the fairness index shown in Figure~\ref{fig:dynamic_fairness}, which indicates that the distributed processing is more equitably spread among the UAVs. Energy efficiency, an important consideration in UAV operations, is also enhanced by congestion-aware early-exit mechanism, as evidenced by the lower average energy per task depicted in Figure~\ref{fig:dynamic_energy}. Ultimately, the overall figure of merit (Figure~\ref{fig:dynamic_merit}) encapsulates these improvements, demonstrating that congestion-aware early-exit mechanism not only enhances individual performance metrics but also provides a robust and scalable solution for collaborative task offloading in large-scale UAV swarms. This approach is particularly valuable in environments where network conditions are highly dynamic, and it paves the way for future enhancements in adaptive task allocation strategies.

\section{Conclusion}
\label{sec:conclusion} 
In this paper, we proposed a fully distributed, diffusive metric-based solution to support split computing in large-scale UAV swarms. By introducing the concept of \textit{aggregated computation capability}, each node can dynamically compute an iterative, local measure that reflects both its own and its neighbors’ computational capacities, without requiring global synchronization or a central coordinator. This approach proved especially suitable for aerial networks with rapidly shifting topologies, heterogeneous hardware resources, and high task generation rates. 

Our simulation results demonstrated that the proposed method effectively alleviates the congestion and bottlenecks that often arise in conventional centralized schemes. By shifting partial inferences (either raw inputs or intermediate layers) toward lightly loaded or computationally advantaged nodes, the swarm as a whole achieves lower latency, improved fairness in task distribution, and better energy efficiency. Moreover, we showed that our solution is robust under a variety of operational conditions, including different UAV speeds, varying mission areas, and fluctuating task arrival rates. It provides a natural mechanism for scalability: as the swarm expands and node-to-node distances change, the diffusive updates and local decisions continue to steer computational tasks toward favorable regions of the network. 
\color{black}
For healthy operation in the field, the swarm needs (i) lightweight runtime monitoring of per-node load and link quality, (ii) clear mission-time rules for tuning $\gamma$, $\tau_{\text{med}}$, and $\tau_{\text{high}}$ across heterogeneous UAVs, and (iii) an explicit accuracy policy for early-exit.
\color{black}

The proposed approach opens several avenues for future work. One potential extension involves incorporating advanced mobility models and real flight data to capture even more realistic UAV trajectories. Additionally, while we primarily focused on split learning at the inference level, further exploration could integrate this method with federated training, enabling on-the-fly updates to model parameters in addition to distributing intermediate layers. Another intriguing direction involves dynamically adjusting the structure of the neural network layers themselves, employing model compression or pruning techniques, to further reduce communication overhead when bandwidth constraints become severe. We expect that, taken together, these refinements will push the frontier of distributed intelligence in swarms, enabling UAV networks of unprecedented scale and complexity to function reliably in even the most challenging environments.

\section*{Acknowledgment}

This work is supported by Istanbul Technical University, Department of Scientific Research Projects (ITU-BAP, 45375), and was partially supported by the PRIN 2022 project RAIN4C: ``Reliable Aerial and satellIte Networks: joint Communication, Computation, Caching for Critical scenarios'' (Id: 20227N3SPN, CUP: J53D23007020001).

% \bibliographystyle{IEEEtran}
% \bibliography{IEEEabrv,bibliography}
\bibliographystyle{elsarticle-num-names} 
\bibliography{bibliography}

\end{document}